\documentclass[aps,prx,twocolumn,longbibliography,superscriptaddress,showpacs]{revtex4-2}
\usepackage{float}
\usepackage{caption}
\usepackage{ragged2e} 
\usepackage{subcaption}
\usepackage{graphicx} 
\graphicspath{{figures}}
\usepackage{color}
\usepackage{xcolor}
\usepackage{bm}        
\usepackage{braket}
\usepackage{amssymb}   
\usepackage{epstopdf}
\usepackage{xfrac}
\usepackage{cancel}
\usepackage{soul}
\usepackage{amsmath}
\usepackage{amsthm}
\usepackage{amsfonts}
\usepackage{bbm}
\definecolor{darkblue}{rgb}{0.1,0.2,0.6}
\definecolor{darkred}{rgb}{0.8,0.1,0.2}
\definecolor{darkgreen}{rgb}{0.31,0.62,0.24}
\definecolor{OliveGreen}{cmyk}{0.64, 0, 0.95, 0.40}
\definecolor{purple}{rgb}{0.5,0.1,0.5}

\usepackage[colorlinks,citecolor=darkblue,linkcolor=darkblue,urlcolor=darkblue]{hyperref} 
\usepackage{enumerate}
\usepackage{setspace}
\usepackage{url}  
\usepackage{mathrsfs}

\usepackage[percent]{overpic}

\newcommand{\Tr}{\text{Tr}}

\newcommand{\hgamma}{\hat \gamma}
\def\i{\textbf{i}}
\def\m{\mathfrak{m}}

\def\IsGS{|\Omega\rangle}
\def\Ceff{c_{\rm eff}}

\newcommand{\cmjadd}[1]{
}
\newcommand{\cmjCom}[1]{
}
\newcommand{\zyCom}[1]{
}
\newcommand{\zyadd}[1]{
}
\newcommand{\dmadd}[1]{
}
\newcommand{\dmCom}[1]{
}

\begin{document}

\title{Entanglement in a one-dimensional critical state after measurements}

\author{Zhou Yang}
\author{Dan Mao}
\author{Chao-Ming Jian}
 \affiliation{Department of Physics, Cornell University, Ithaca, New York 14853, USA}

\date{\today}

\begin{abstract}

The entanglement entropy (EE) of the ground state of a one-dimensional Hamiltonian at criticality has a universal logarithmic scaling with a prefactor given by the central charge $c$ of the underlying 1+1d conformal field theory. When the system is probed by measurements, the entanglement in the critical ground state is inevitably affected due to wavefunction collapse. In this paper, we study the effect of weak measurements on the entanglement scaling in the ground state of the one-dimensional critical transverse-field Ising model. For the measurements of the spins along their transverse spin axis, we identify interesting post-measurement states associated with spatially uniform measurement outcomes. The EE in these states still satisfies the logarithmic scaling but with an alternative prefactor given by the effective central charge $\Ceff$. We derive the analytical expression of $\Ceff$ as a function of the measurement strength. With both numerical simulations and analytical studies, we show that for the EE averaged over all post-measurement states based on their Born-rule probabilities, the effective central charge is independent of the measurement strength, contrary to the usual expectation that local and non-overlapping measurements reduce the entanglement in the system. We also examine the behavior of the average EE under (biased) forced measurements where the measurement outcomes are sampled with a pre-determined probability distribution without inter-site correlations. In particular, we find an optimal probability distribution that can serve as a mean-field approximation to the Born-rule probabilities and lead to the same $\Ceff$ behavior. The effects of the measurements along the longitudinal spin axis and the post-measurement correlation functions are also discussed.

\end{abstract}

\maketitle

\section{Introduction}

For a quantum system, measurements can affect its state in a highly non-trivial manner. Measurement is an intrinsically non-unitary operation due to the wavefunction collapse in the process. Therefore, its effects on quantum states fundamentally differ from any unitary operations. One can exploit the peculiar effect of measurement to either generate or reduce the entanglement in a quantum system. For example, the entanglement between the Einstein-Podolsky-Rosen pair is eliminated once one partner is measured. On the other hand, one can also simultaneously perform a set of commuting measurements on overlapping observables on an array of un-entangled qubits to create long-range-entangled states with topological order, such as the toric code \cite{Kitaev1997,Kitaev2003, LevinWenModel}, which are states that cannot be created by the local unitary circuit with a depth independent of the system size \cite{ZengWen2015}. Moreover, starting with many-body quantum states with certain entanglement structures (such as the cluster states), general quantum information processing can be carried out by sequences of measurements, following the scheme of measurement-based quantum computation \cite{Raussendorf2001OneWay,Gottesman1999,Raussendorf2003,Hans2001}. From a quantum matter perspective, the ground states of quantum many-body systems (with local Hamiltonians) provide a large class of states with rich entanglement properties \cite{Cardy2004,LevinWenTopoEE,MyersSinha2010,MyersSinha2011,CasiniHuerta2007,CasiniHuerta2012,LiHaldane2008,MetlitskiGrover2011}. Any form of probe/measurement will inevitably alter the many-body wavefunctions and, hence, change the entanglement structure in the original ground states. In general, it is interesting to understand how measurements affect the entanglement in quantum many-body systems.

\cmjadd{Among the ground states of local quantum Hamiltonian systems, critical ground states are among the most entangled.} In particular, for a one-dimensional critical system, the ground-state von Neumann entanglement entropy (EE) on a subsystem of length $l$ has a \cmjadd{universal} logarithmic scaling $\sim \frac{c}{3}\log l$ 
\cmjadd{where $c$ is the central charge of the conformal field theory (CFT) that describes the low-energy physics of the system \cite{calabrese2009}}. Examples of non-trivial effects of measurements on a one-dimensional critical ground state were presented in interesting recent works Ref. \onlinecite{LinZouSanHsieh2022} and \onlinecite{Garratt2022}. For instance, the latter work showed that the correlation functions in a (one-dimensional) Tomonaga-Luttinger liquid can experience transitions when the state is measured. 
More generally, the problem of a one-dimensional critical ground state subject to measurements is naturally mapped to the problem of boundary\cmjadd{/defect} CFTs (or critical systems with impurities/defects) \cite{Cardy2004,AFFLECK1990,AFFLECK1991,KaneFisher1992}, which is a subject that has been extensively studied. Similar ideas generalize naturally to higher dimensions. In another recent work Ref. \onlinecite{LeeJianXu2023}, phase transitions have also been found in two-dimensional critical states subject to weak measurements.

In this work, we are interested in the entanglement structure, the EE in particular, in the one-dimensional critical system after measurements are performed on the ground state.

where $\lambda=1$ indicates the standard projective measurements, and the measurement is reduced to a trivial action on the state in the limit $\lambda = 0$. After the measurement, each set of measurement outcomes corresponds to a particular post-measurement (pure) state due to the wavefunction collapse. The post-measurement states for different sets of measurement outcomes form an ensemble. It is interesting to explore different ways to sample this ensemble and study the ${\it universal}$ behavior of the EE in the post-measurement states. \cmjadd{Later, we will see that the measurements along the $x$- or $z$-axis have different properties under the Ising $\mathbb{Z}_2$ symmetry and lead to distinct classes of universal behaviors in the large system limit. These universal behaviors reflect the properties of the defects CFT at the renormalization group (RG) fixed points, which are independent of the microscopic details of the weak measurements.}

When the spins are independently measured along their $x$-axis, we identify intriguing post-measurement states with spatially uniform measurement outcomes. The EEs in such states still follow the logarithmic scaling  $\frac{\Ceff}{3} \log(l)$ but with an effective central charge $\Ceff$ that decreases continuously and monotonically as the measurement strength $\lambda$ increases. Such behavior results from the fact that the measurement along the $x$-axis induces an exactly marginal perturbation to cTFIM ground state $\IsGS$ in the RG sense. We obtain the exact analytical expression of $\Ceff(\lambda)$ using a combination of a field-theoretic treatment and a mapping to another cTFIM with a defect. We also show analytically that a two-point longitudinal-spin correlation function in these particular post-measurement states exhibits a scaling exponent that varies continuously as the measurement strength changes. \cmjadd{Both the $\Ceff(\lambda)$ and the varying exponents are directly related to the universal properties of an exactly marginal conformal defect in the Ising CFT (with $\lambda$ parametrizing the defect strength)}. When the spins are measured along the $z$-axis, the post-measurement states with uniform measurement outcomes have a vanishing central charge immediately when the measurement strength $\lambda$ is non-zero. As we show later, this behavior results from a relevant perturbation to the cTFIM ground state $\IsGS$ induced by the $z$-axis measurements. \cmjadd{The obtained properties of the uniform post-measurement states can also be interpreted as results in quantum dynamics under non-Hermitian Hamiltonians. }

We also study the average behavior of the EE in the entire ensemble of post-measurement states. For the study of average EE, we focus on the measurements with respect to the $x$-axis of the spins. When the measurements are performed, different measurement outcomes and their associated post-measurement states naturally occur according to the classic Born-rule probability. When physical quantities, including EE, are averaged over the post-measurement states according to Born-rule probability, we refer to this type of measurement as the {\it Born-rule measurement}.
There is another commonly discussed averaging scheme, associated with the so-called {\it forced measurement}, where different measurement outcomes are ``forced" to be sampled with equal probability. One can view the forced measurement as the Born-rule measurement followed by a re-weighting on the ensemble of outcomes and their associated post-measurement states. Although being less natural compared to the Born-rule measurement, in the different context of dynamical quantum systems monitored by measurements, forced measurements were studied in many examples due to their close relations to the physics of random tensor network or disordered statistical mechanics models \cite{Nahum2021AlltoAll,JianYouVassuerLudwig2020MIPT,BaoChoiAltman2020,LiVasseurFisherLudwig2021,JianRTN,VasseurPotterYouLudwigRTNPRB2019}. For the present work, we are interested in the behavior of both the post-measurement EE averaged in accordance with Born-rule measurements and that with forced measurements.  

In the case of Born-rule measurements, we show the average EE can be described by a 1+1d $R$-replica field theory with a measurement-induced perturbation on a one-dimensional defect in the replica limit $R\rightarrow 1$. Using the numerical method based on matrix product states (MPS) (and an alternative method based on a Majorana-fermion representation of the system), we calculate the post-measurement EE averaged with respect to the Born-rule probability. We find that the numerically extracted effective central charge $\Ceff$, which is associated with the logarithmic scaling of the average EE, turns out to be the same as that of the un-measured ground state. This interesting result is contrary to the naive expectation that measurements on local and non-overlapping observables generally reduce the entanglement in the system. We provide an analytical understanding of this behavior of the effective central charge by studying the measurement-induced perturbation \cmjadd{in the CFT and showing that the defect associated with perturbation has a vanishing strength} in the replica limit $R\rightarrow 1$.

In the case of forced measurements, we show the average EE can be captured by the same 1+1d $R$-replica field theory with a measurement-induced perturbation on a one-dimensional defect but in a different replica limit $R\rightarrow 0$. The effective central charge $\Ceff$ extracted from the numerically calculated average EE decreases continuously as the measurement strength increases. Interestingly, one can deform the probability distribution of the forced measurements to set a bias towards one of the two measurement outcomes on every spin. We find an optimal bias under which a mean-field approximation to the Born-rule probability is achieved. With this optimal bias, the numerically calculated effective central charge restores the un-measured value independent of the measurement strength, a feature shared by the average EE with Born-rule measurements. Analytical understandings of such behaviors of the effective central charge under biased and unbiased forced measurements are provided.

The remainder of the paper is organized as follows. In Sec. \ref{Sec:Measure_cTFIM}. we introduce cTFIM and the weak measurements on its ground state that we consider in this work. In Sec. \ref{sec:post-selection_EE}, we study the universal properties, especially the effective central charges, of the post-measurement states with uniform measurement outcomes. Both exact analytical results and their numerical verification will be presented.
In Sec. \ref{sec:born}, we discuss the average EE with Born-rule measurement using both analytical and numerical methods. In Sec. \ref{sec:force}, we present our results on the average EE with forced measurements and biased forced measurements obtained from both analytical and numerical approaches. In Sec. \ref{sec:summary}, we provide a summary of the results of this paper and discuss several interesting questions for future investigation.

\section{Measuring the ground state of the critical transverse field Ising model}
\label{Sec:Measure_cTFIM}

We consider the ground state $|\Omega\rangle$ of the cTFIM on a spin chain whose Hamiltonian is given by
\begin{equation}
H = -\sum_j (\sigma^z_j \sigma^z_{j+1} + \sigma^x_j)
\label{eq:TFIM}
\end{equation}
where $\sigma^{x,z}_j$ are the Pauli operators on the $j$th site of the spin chain. \cmjadd{The Hamitonian $H$ and its ground state $\IsGS$ preserves the global Ising $\mathbb{Z}_2$ symmetry: $\sigma^z_j \rightarrow -\sigma_j^z$.}

Let us first introduce the types of measurements on the ground state $|\Omega\rangle$ we will study in this paper. We will primarily focus on weak measurements of spins in the system along their $x$-axis. We will also discuss the effect of the $z$-axis weak measurement as well. \cmjadd{As explained below, these two types of measurements have different properties under the Ising $\mathbb{Z}_2$ symmetry and generate perturbations to the underlying Ising CFT through different primary operators.}

The weak measurement we consider is a softened version of the standard projective measurement \cmjadd{(which can be implemented by using ancilla qubits and projective measurement \cite{wiseman_milburn_2009})}. For a given site $j$, the $x$-axis weak measurement is defined using the Kraus-operator set $\{K^x_{j,\m_j}\}_{\m_j=\pm}$ with 
\begin{align}
    K^x_{j,\pm} = \frac{1\pm\lambda\sigma^x_j}{\sqrt{2(1+\lambda^2)}},
    \label{eq:KrausX}
\end{align}
where $\lambda \in [0,1]$ is the parameter that controls the strength of the measurement. For this $x$-axis weak measurement, there are two possible measurement outcomes $\m_j = \pm $. Depending on the measurement outcome $\m_j$, an incoming state $|\psi\rangle$ evolves/collapses into one of the two post-measurement states according to 
\begin{align}
    |\psi\rangle \rightarrow \frac{K^x_{j,\m_j} |\psi\rangle }{||  K^x_{j,\m_j} |\psi\rangle || },
    \label{eq:wavefunction_collapse}
\end{align}
This evolution is commonly referred to as the {\it quantum trajectory} associated with the measurement outcome $\m_j$. 

The two quantum trajectories with $\m_j = \pm $ occur with the classic Born-rule probability
\begin{align}
    p^x(\m_j) &= \langle \psi | (K^x_{j,\m_j})^\dag K^x_{j,\m_j}  |\psi \rangle 
    \nonumber \\
    &= \frac{1}{2(1+\lambda^2)}\Big( 1+\lambda^2 + 2\m_j \lambda  \langle \psi | \sigma_j^x |\psi \rangle  \Big).
    \label{eq:Born-ruleX}
\end{align}
The Kraus operators $\{K^x_{j,\m_j}\}_{\m_j=\pm}$ satisfy the condition for a positive operator-valued measure (POVM):
\begin{align}
    \sum_{\m_j = \pm } (K^x_{j,\m_j})^\dag K^x_{j,\m_j} = \openone,
\end{align}
which guarantees the normalization of total Born-rule probability given in  Eq. \eqref{eq:Born-ruleX}, i.e. $\sum_{\m_j=\pm}p^x(\m_j) = 1$, for any incoming state. (See, for example, Ref. \onlinecite{NielsenChuang2010} for a general discussion of the Kraus-operator formalism and the POVM condition).

When $\lambda = 1$, the weak measurement defined by the Kraus-operator set $\{K^x_{j,\m_j}\}_{\m_j=\pm}$ reduces to the standard projective measurement with respect to observable $\sigma_j^x$. And the Kraus operators reduce to $\frac{1 + \m_j \sigma_j^x}{2}$ which are the projection operators to the eigenstates of $\sigma_j^x$.  For $0<\lambda < 1$, the measurement has a ``lower resolution" in resolving the spin state along the $x$-axis
and the Kraus operators in Eq. \eqref{eq:KrausX} become softened versions of the projection operators. In the limit $\lambda=0$, the Kraus operators are reduced to the identity operators (up to a multiplicative constant). The measurement does not provide any information on the state of the system. Hence, there is effectively no measurement performed.

Similarly, the $z$-axis weak measurement is defined by the Kraus-operator set $\{K^z_{j,\m_j}\}_{\m_j=\pm}$ with 
\begin{align}
    K^z_{j,\pm} = \frac{1\pm\lambda\sigma^z_j}{\sqrt{2(1+\lambda^2)}},
    \label{eq:KrausZ}
\end{align}
which also satisfies the POVM condition:
\begin{align}
    \sum_{\m_j = \pm } (K^z_{j,\m_j})^\dag K^z_{j,\m_j} = \openone.
\end{align}
For this $z$-axis weak measurement, the definition of quantum trajectories and the expressions of the Born-rule probabilities for different measurement outcomes $\m_j=\pm$ parallel the case of the $x$-axis weak measurement discussed above. 

Consider performing a weak measurement along the spin axis $v$ on every spin in the ground state $|\Omega\rangle$ of the cTFIM. There are two cases given by $v=x$ and $v=z$. On a spin chain of length $L$, the set of measurement outcomes from the measurements on every site is denoted as $\{\m_j\}_{j=1,2,...,L}$. In the quantum trajectory associated with $\{\m_j\}$, the post-measurement state is given by
\begin{align}
    \ket{\Psi^v_{\{\m_j\}} } = \frac{\prod_j K^v_{j,\m_j}\ket{\Omega} }{|| \prod_j K^v_{j,\m_j}\ket{\Omega} ||}.
\end{align}
\cmjadd{When the measurement axis is $v=x$, the post-measurement states in all quantum trajectories preserve the Ising $\mathbb{Z}_2$ symmetry. When $v=z$, the Ising $\mathbb{Z}_2$ symmetry is broken in the post-measurement states.}

In this paper, we are interested in the entanglement properties, especially the EE, of the post-measurement states $\ket{\Psi^v_{\{\m_j\}} }$.
Before any measurement, on a spin chain of length $L$, the half-system EE follows the logarithmic scaling
\begin{equation}
S_{\IsGS} (L/2) = \frac{c}{6}\log(L) + O(1)
\label{eq:EE_log_scaling}
\end{equation}
where $c=1/2$ is the central charge of 1+1d Ising conformal field theory (CFT) that governs the low-energy physics of cTFIM. Note that Eq. \eqref{eq:EE_log_scaling} applies to a one-dimensional system with an open boundary condition where the half system refers to an interval of length $L/2$ starting from the boundary of the spin chain. This expression of EE, especially the prefactor $c/6$, can be derived by introducing the twist fields in Ising CFT \cite{calabrese2009}. The previously mentioned universal EE scaling $\frac{c}{3}\log(l)$, which concerns a subsystem of length $l$ away from the boundary of the system, has exactly the same origin as Eq. \eqref{eq:EE_log_scaling}. The factor of 2 difference between the two EE scaling expressions is related to the difference in the number of boundary points of the subsystems.

After the weak measurements on every site of the system, the ensemble of different possible sets of measurement outcomes gives rise to an ensemble of different post-measurement states $\{\ket{\Psi^v_{\{\m_j\}}} \}$. \cmjadd{These post-measurement states can be viewed as the Ising CFT ground state perturbed by primary operators that depend on $v$, which we will elaborate in the following sections.} Depending on the way the post-measurement states are sampled, our study is divided into three parts. 

In the first part, we focus on the specific post-measurement states $\ket{\Psi^v_{\{\m_j = +\}} }$ and $\ket{\Psi^v_{\{\m_j = -\}} }$ associated with the spatially uniform measurement outcomes, i.e. $\m_j = +$ for all sites or $\m_j = -$ for all sites. 

The second part concerns the case of Born-rule measurement, where we study the average EE over all post-measurement states weighted by their associated Born-rule probability
\begin{equation}
    p(\{\m_j\}) = \Big\langle\Omega \Big| \prod_j K^{v\dagger}_{j,\m_j} K^v_{ j,\m_j}\Big|\Omega\Big\rangle\label{eq:Born-Rule_Chain}.
\end{equation}

In the third part, we consider the case of forced measurements where all the post-measurement states $\ket{\Psi^v_{\{\m_j \}} }$ are ``forced" to be sampled with equal probability. We will also introduce a biased version of the forced measurement where we can put in a tunable bias in the statistical weight for $\m_j = +$ and $\m_j = -$ on each site.

\section{Exact result for uniform measurement outcomes}
\label{sec:post-selection_EE}

\begin{figure*}[tb]
\centering
    \captionsetup{justification = RaggedRight}
    \subfloat[\label{fig:Perturbed_Ising_Twists1}]{\includegraphics[width = .40\linewidth]{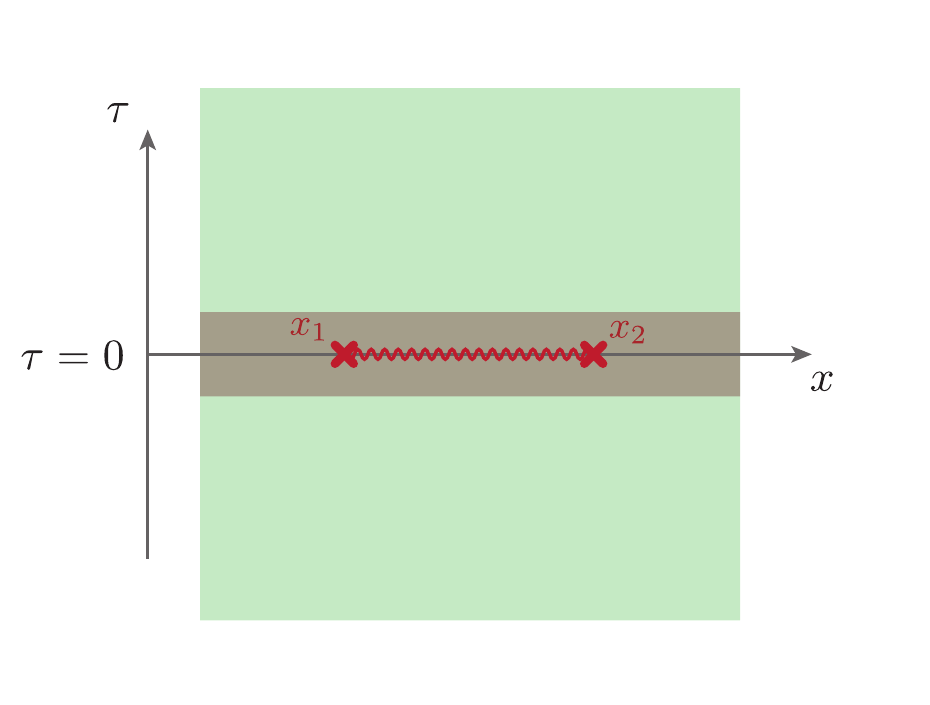}}
    \hspace{0.1\linewidth}
    \subfloat[\label{fig:Perturbed_Ising_Twists2}]{\includegraphics[width = .40\linewidth]{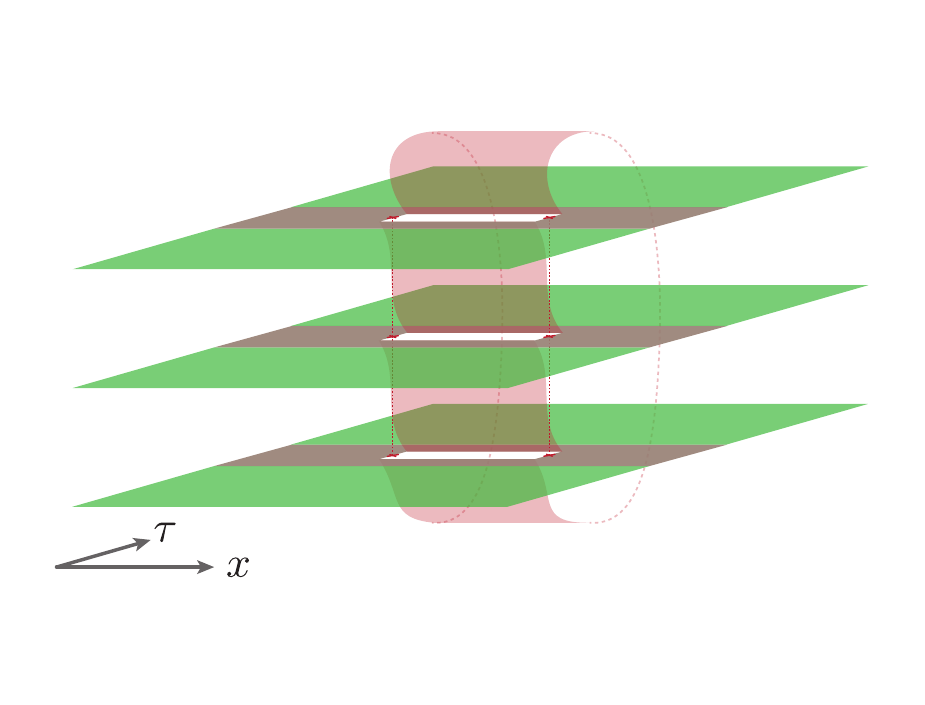}}
    \caption{(a) A pictorial representation of the path integral for the post-measurement state $\ket{\Psi^v_+}$. The green region is described by the 1+1d Ising CFT. The brown region denotes the region where the perturbation induced by the measurements is turned on. To calculate the EE on the interval $A = [x_1, x_2]$ in the post-measurement state $\ket{\Psi^v_+}$, one should consider the $R$-replica version of this path integral and 
    the insertion of the twist fields $\mathcal{T}_R$ at endpoints of the interval (indicated by the red crosses)       
    (b) $R$-replica path integral representation of $\Tr \left(\rho_{[x_1,x_2]}^R\right)$ for the reduced density matrix $\rho_{[x_1,x_2]}$ of the state $\ket{\Psi^v_+}$ on the interval $[x_1,x_2]$. The insertion of the twist fields $\mathcal{T}_R$ and $\tilde{\mathcal{T}}_R$ introduces an $R$-fold branch cut in the path integral. }
    \label{fig:Perturbed_Ising_Twists} 
\end{figure*}

In this section, we focus on the post-measurement states $\ket{\Psi^v_{\{\m_j = +\}} }$ and $\ket{\Psi^v_{\{\m_j = -\}} }$
with spatially uniform measurement outcomes, i.e. $\m_j = +$ for all sites or $\m_j = -$ for all sites. For simplicity, we denote them as $\ket{\Psi^v_\pm}$ respectively. Most interestingly, we show that when the weak measurements are performed with respect to the spin axis $v = x$, the half-system EEs of the post-measurement states $\ket{\Psi^x_\pm}$ follow a similar logarithmic scaling as in Eq. \eqref{eq:EE_log_scaling}. However, the central charge $c=1/2$ is replaced by an effective central charge $\Ceff$ that decreases monotonically as the measurement strength $\lambda$ increases. The analytical expression of $\Ceff$ as a function of $\lambda$ is derived using a combination of a field-theoretic treatment and a mapping to an Ising chain with a defect. 
We also show that the scaling exponents associated with the correlation function $\langle \sigma_j^z\sigma_{j'}^z\rangle$ in the post-measurement states $\ket{\Psi^x_\pm}$ can be deformed continuously as a function of $\lambda$. Exact expressions of these exponents are obtained. 

In contrast, when the weak measurements are performed with respect to the spin axis $v = z$, we show that the EEs of the post-measurement states $\ket{\Psi^z_\pm }$ do not have any logarithmic dependence on the system size $L$ for any non-vanishing $\lambda>0$. In other words, the effective central charge $\Ceff$ drops from $c = 1/2$ at $\lambda = 0$ (where there is essentially no measurement) to $\Ceff = 0$ for any non-zero $\lambda$.

We begin our analysis of the post-measurement states $\ket{\Psi^v_\pm}$ by providing a field-theoretic interpretation of them. We will use the state $\ket{\Psi^v_+}$ as an example and comment on the difference between $\ket{\Psi^v_+}$ and $\ket{\Psi^v_-}$ when necessary. The post-measurement state $\ket{\Psi^v_+}$ can be rewritten, up to a normalization factor, as
\begin{align}
    \ket{\Psi^v_+} \propto e^{\beta \sum_j \sigma^v_j} \IsGS,
\end{align}
where $\beta = {\rm arctanh}(\lambda)$. 
In the continuum description of the infinite-length chain ($L \rightarrow \infty$), the cTFIM ground state $\IsGS$ can be viewed as the result of the Euclidean path integral of the 1+1d Ising CFT on a half-space with the spatial coordinate $x \in (-\infty, \infty)$ and the imaginary time $\tau \in (-\infty, 0 ]$. $\IsGS$ occurs at the $\tau = 0$ time slice. Therefore, any observable $\bra{\Psi^v_+} \mathcal{O} \ket{\Psi^v_+}$ on the post-measurement state $\ket{\Psi^v_+}$ can be captured by a path integral in a full Euclidean plane $(x,\tau) \in \mathbb{R}^2$ with an action:
\begin{align}
    \mathcal{S}_{{\rm ps},+}^v = \mathcal{S}_{\rm Ising} + \delta \mathcal{S}_{{\rm ps},+}^v 
    \label{eq:S_post_selection}
\end{align}
where $\mathcal{S}_{\rm Ising} $ is the action for the 1+1d Ising CFT and the measurment-induced perturbation $\delta \mathcal{S}_{{\rm ps},+}^v$ is given by
\begin{align}
    \delta \mathcal{S}_{{\rm ps},+}^v \equiv -\Tilde{\beta} \int d\tau dx ~ \delta(\tau)  \phi^v(x,\tau),
    \label{eq:S_perturb_post_selection}
\end{align}
\cmjadd{which can be viewed as a one-dimensional defect in the 1+1d Ising CFT.} When the measurement axis is $v=x$, the field $\phi^{v=x} = \varepsilon$ is the energy field of the Ising CFT, while $\phi^{v=z} = s$ is the spin field (or the order parameter) of the Ising CFT when the measurement axis is $v=z$. The coupling constant $\Tilde{\beta}$ is related to the measurement strength $\lambda$. The exact relation between $\Tilde{\beta}$ and $\lambda$ is unimportant and also non-universal.

Eq. \eqref{eq:S_post_selection} describes the Ising CFT perturbed by the field $\phi^v$ turned on only along the $\tau = 0$ slice. Replacing the perturbation $\delta \mathcal{S}_{{\rm ps},+}^v$ by $ -\int_{-\Tilde{\beta}_0/2}^{\Tilde{\beta}_0/2} d\tau \int^{\infty}_{-\infty} dx  \phi^v(x, \tau) $ within a finite time window $\tau \in [-\Tilde{\beta}_0/2,\Tilde{\beta}_0/2]$ does not change the infrared behavior of the path integral Eq. \eqref{eq:S_post_selection}.  We can pictorially represent this path integral of the perturbed Ising CFT as Fig. \ref{fig:Perturbed_Ising_Twists1} where the green regions represent the unperturbed Ising CFT path integral and the brown region around $\tau = 0$ is the region where the perturbation is turned on. It is interesting to point out that if we fold spacetime along the $\tau=0$ axis, we can view this path integral as that of a boundary CFT where the bulk is given by a double-layer Ising CFT. However, we will not directly use this picture of folded spacetime in the following discussion. 

From the renormalization group (RG) perspective, the scaling dimension of the coupling $\tilde{\beta}$ is given by
\begin{align}
    \Delta_{\tilde{\beta}} = 1 -  \Delta_{\phi^v},
    \label{eq:post_measurement_scaling_dim}
\end{align}
where $\Delta_{\phi^v}$ is the scaling dimension of the field $\phi^v$. For the measurement axis $v=x$, $\Delta_{\phi^x} = \Delta_{\epsilon} =1 $ is the scaling dimension of the energy field of the Ising CFT. Hence, $\Delta_{\tilde{\beta}} = 0$ which implies that the perturbation is marginal in the sense of RG. For the measurement axis $v=z$, $\Delta_{\phi^z} = \Delta_{s} =1/8 $ is the scaling dimension of the spin field. In this case, $\Delta_{\tilde{\beta}} = 7/8$ indicating a relevant perturbation.

Formally we can calculate the EE in the state $|\Psi_+^v\rangle$ by introducing the twist fields in the path integral described by Eq. \eqref{eq:S_post_selection}. This calculation generalizes the well-known derivation of Eq. \eqref{eq:EE_log_scaling} of the EE in the ground state of an (unperturbed) 1+1d CFT using the twist-field correlation functions \cite{calabrese2009}. For the EE of an interval $[x_1, x_2]$ in the state $|\Psi_+^v\rangle$ (see Fig. \ref{fig:Perturbed_Ising_Twists1}), we need to first introduce $R$ replicas of the path integral governed by Eq. \eqref{eq:S_post_selection}. Then, we introduce the twist field $\mathcal{T}_R$ and its conjugate $\tilde{\mathcal{T}}_R$ at $(x_1,\tau = 0)$ and $(x_2,\tau = 0)$ that serve as the endpoints of an $R$-fold branch cut along the interval $[x_1,x_2]$ in this $R$-replica path integral (see Fig. \ref{fig:Perturbed_Ising_Twists2}). The EE of the interval $[x_1, x_2]$ in the post-measurement state $|\Psi_+^v\rangle$ can be expressed as 
\begin{align}
    &S_{|\Psi_+^v\rangle}([x_1,x_2]) = -\Tr \left( \rho_{[x_1,x_2]}\log \rho_{[x_1,x_2]} \right)
    \nonumber \\
    & ~~= \lim_{R\rightarrow 1} \frac{1}{1-R}\log \langle \mathcal{T}_R (x_1, \tau =0)  \ \tilde{\mathcal{T}}_R (x_2, \tau =0) \rangle, 
\end{align}
where the $\rho_{[x_1,x_2]}$ is the reduced density matrix on the interval $[x_1, x_2]$.
To be more explicit about the second line, $\langle \mathcal{T}_R (x_1, \tau =0) \ \tilde{\mathcal{T}}_R (x_2, \tau =0) \rangle$ is defined as the ratio of $R$-replica path integral with and without the twist fields, that is with and without the $R$-fold branch cut. In terms of the reduced density matrices $\rho_{[x_1,x_2]}$, the twist field two-point function is equivalent to $\Tr \left(\rho_{[x_1,x_2]}^R\right)/\left(\Tr \rho_{[x_1,x_2]}\right)^R$.

\begin{figure}[tb]
 \centering
\captionsetup{justification = RaggedRight}
\includegraphics[width = 0.9\linewidth]{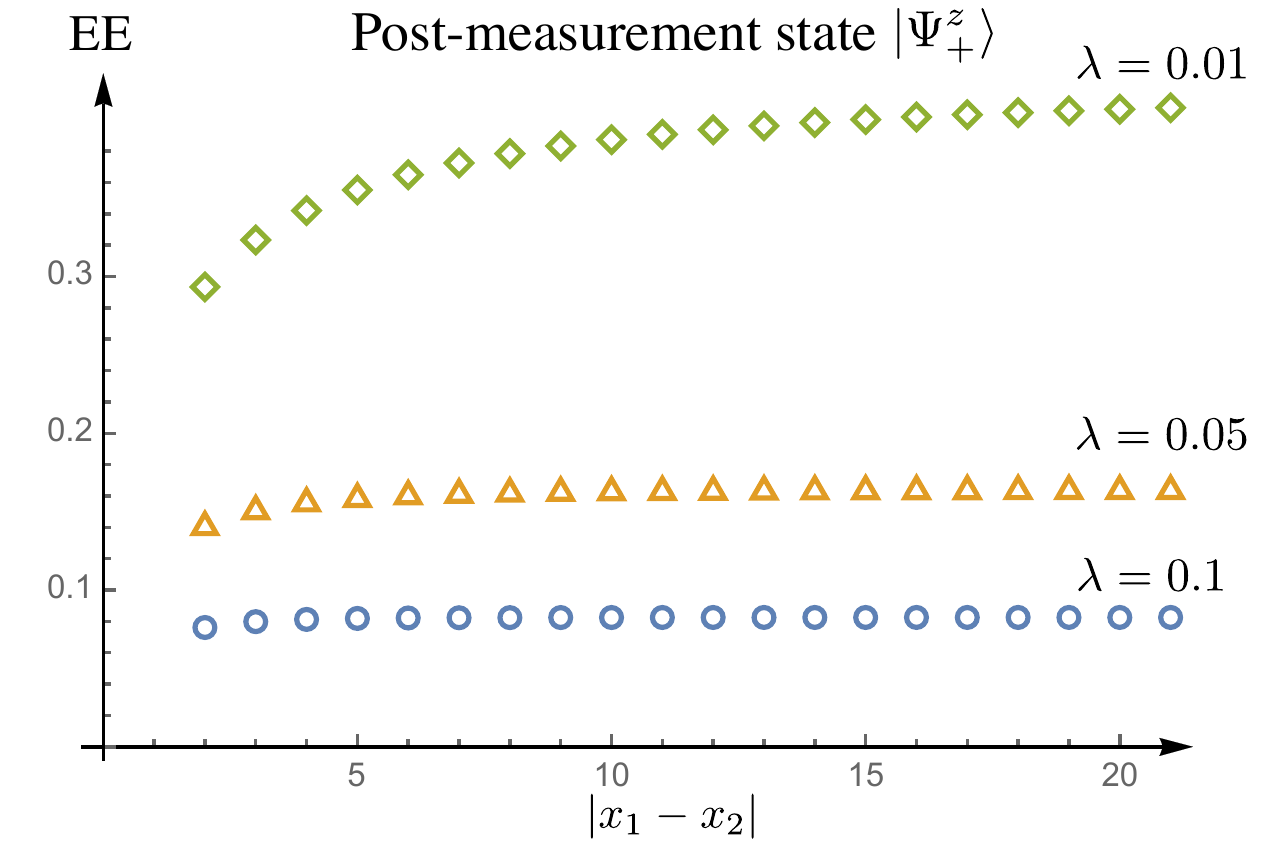}
\caption{For the post-measurement state $|\Psi_+^z\rangle$ obtained from measuring the cTFIM ground state $\IsGS$ along the spin axis $v=z$, the EE $S_{|\Psi_+^z\rangle}([x_1,x_2])$ on an interval $[x_1,x_2]$ is plotted as a function of $|x_1-x_2|$ for different measurement strength $\lambda= 0.1, 0.05, 0.01$. $S_{|\Psi_+^z\rangle}([x_1,x_2])$ saturates as the interval $|x_1-x_2|$ grows big.
}\label{fig:sigmaz1}
\end{figure}

When the weak measurements are performed along the spin axis $v=z$, the perturbation $\delta \mathcal{S}_{{\rm ps},+}^z$ is relevant under RG. Such a perturbation effectively cuts the $(x,\tau)$ plane along the $\tau = 0$ line (the brown region in Fig. \ref{fig:Perturbed_Ising_Twists}) into two decoupled halves with $\tau>0$ and $\tau<0$. In this case, the $R$-fold branch cut associated with twist fields $\mathcal{T}_R $ and $\tilde{\mathcal{T}}_R $, which are located right in the middle of the cut, has no effect in IR behavior of the path integral. Hence, we expect that the correlation $\langle \mathcal{T}_R (x_1, \tau =0)  \ \tilde{\mathcal{T}}_R (x_2, \tau =0) \rangle$ saturates, in the limit of a large separation between $x_1$ and $x_2$, to a finite value. 
Therefore, the EE $S_{|\Psi_+^z\rangle}([x_1,x_2]) $ should saturate to an $O(1)$ value for a large interval $[x_1, x_2]$. In other words, in the case of the $z$-axis weak measurements, the EE of the post-measurement state $|\Psi_+^z\rangle$ has a vanishing effective central charge, i.e. $\Ceff = 0$, for any non-vanishing value of the measurement strength, i.e, $\lambda>0$. This conclusion is confirmed by the numerical simulation shown in Fig. \ref{fig:sigmaz1}. The EE of the post-measurement state $|\Psi_+^z\rangle$ is calculated using the numerical method based on matrix product states (MPS). We prepare $|\Psi_+^z\rangle$ by starting with an infinite MPS that simulates the ground state of cTFIM and applying an infinite product of $e^{\lambda \sigma_j^z}$ on each site. We then calculate the EE between the interval $[x_1,x_2]$ and the rest of the chain. (See App. \ref{app:dmrg} for more details of the MPS simulation.) The post-measurement state $|\Psi_-^z\rangle$ shares exactly the same behavior as $|\Psi_+^z\rangle$ as they are related to each other by a global symmetry generated by $\prod_j \sigma_j^x$.

\begin{figure}
\centering
\captionsetup{justification = RaggedRight}
    \includegraphics[width = 0.45 \textwidth]{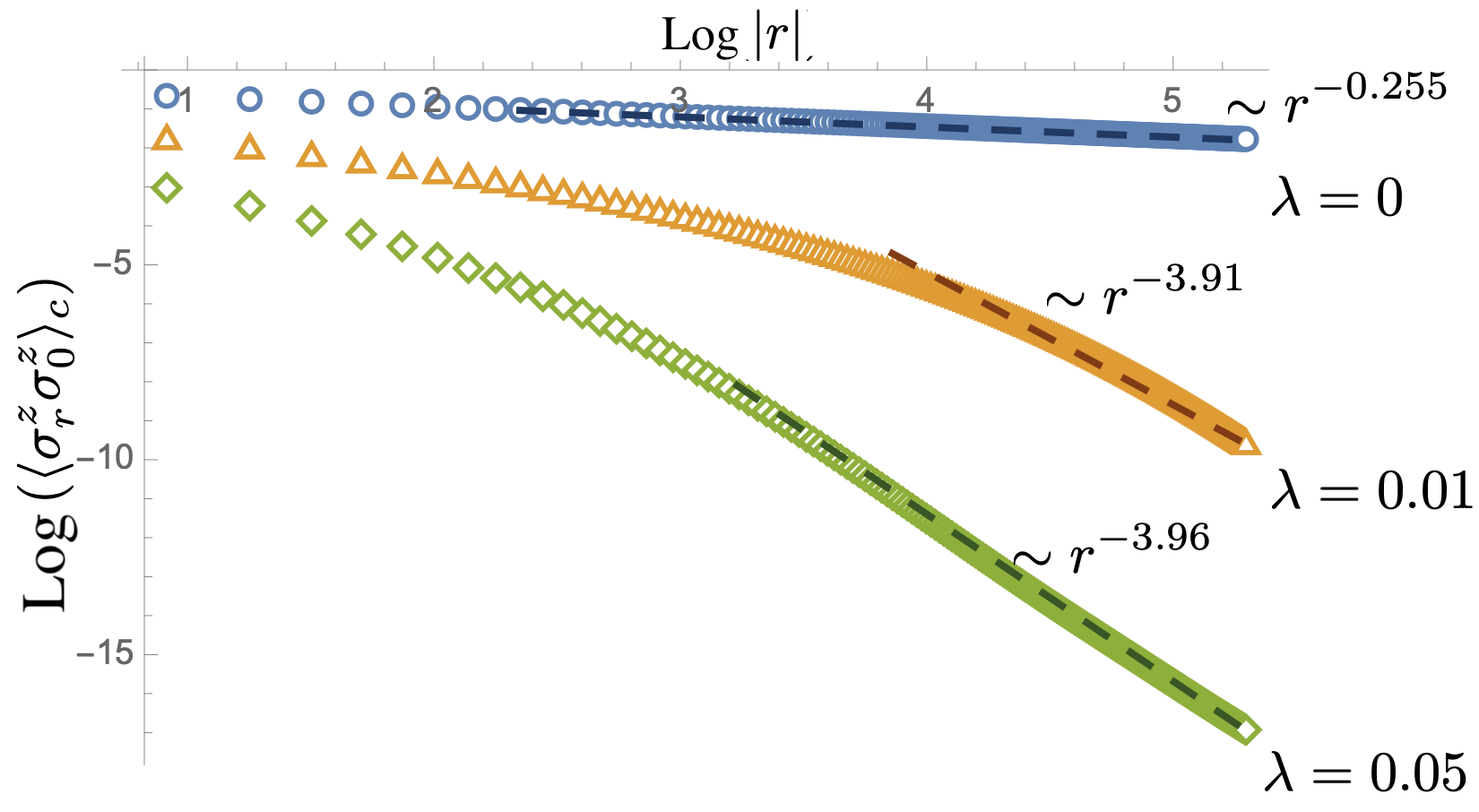}
\caption{Connected $\sigma^z$ correlators after Z-basis measurement for uniform measurement outcome.  $\lambda$ is the strength of the weak measurement. $r$ is the separation between the two operators. The exponents are extrapolated using the correlators at large separation. When $\lambda = 0$, there is no measurement, and we expect the exponent to be $-1/4$ (based on the scaling dimension of the spin field in the Ising CFT). For $\lambda > 0$, the exponents become $\sim -4$, consistent with the extraordinary boundary of the Ising CFT. The simulation is done with bond dimension $65$ for a chain with $1000$ sites. Increasing bond dimension and system size gives convergent results.}
    \label{fig:corr_after_Z}
\end{figure}

\cmjadd{Note that even though $\delta \mathcal{S}_{{\rm ps},+}^z$ is a relevant perturbation that results in a vanishing effective central charge. The post-measurement state $|\Psi_+^z\rangle$ still exhibits non-trivial power-law correlation functions such as $\langle \Psi_+^z|\sigma^z_j \sigma^z_i |\Psi_+^z\rangle_c \sim \frac{1}{|i-j|^4}$, which is expected from the boundary Ising CFT (with the extraordinary boundary) \cite{cardy1984conformal}. Here, the subscript $c$ indicates the connected correlation. We confirm this expectation using the numerical calculation of $\langle \Psi_+^z|\sigma^z_j \sigma^z_i |\Psi_+^z\rangle_c$ in the MPS representation, which is shown in Fig. \ref{fig:corr_after_Z}.}


Now, we consider the weak measurements performed along the spin axis $v=x$. For small $\lambda$, it is easy to show that the two states $|\Psi_+^x\rangle$ and $|\Psi_-^x\rangle$ correspond to the most and the least probable amongst all possible measurement outcomes $\{\m_j\}$. For $|\Psi_+^x\rangle$, 
the \cmjadd{one-dimensional defect caused by the perturbation} $\delta \mathcal{S}_{{\rm ps},+}^x$ is marginal as explained under Eq. \eqref{eq:post_measurement_scaling_dim}. \cmjadd{$\lambda$ can be viewed as the parameter that controls the strength of the defect.} In fact, this defect is exactly marginal, which is evident when we reformulate the 1+1d Ising CFT as the 1+1d free massless Majorana-fermion CFT. The perturbation $\delta \mathcal{S}_{{\rm ps},+}^x$ can be rewritten as the  Majorana-fermion mass term localized along the $\tau = 0$ time slice. The perturbed theory of Majorana fermions remains non-interacting, and there is no RG flow for the localized mass term. Therefore, $\delta \mathcal{S}_{{\rm ps},+}^x$ is exactly marginal. As a consequence, on a chain of length $L$, the half-system EE of the post-measurement state $|\Psi_{+}^x\rangle$ follows a logarithmic scaling 
\begin{align}
    S_{|\Psi_{+}^x\rangle} (L/2) = \frac{\Ceff}{6}\log(L) + O(1),
    \label{eq:EE_log_scaling_x_measured}
\end{align}
which is similar to Eq. \eqref{eq:EE_log_scaling} for the un-measured cTFIM ground state $\IsGS$ but with an effective central charge 
\begin{align}
     \Ceff = 
\frac{-3}{\pi^2}\Big\{\big[(1+s)\log(1+s)+(1-s)\log(1-s)\big]\log(s) \nonumber \\
     \hspace{1cm}  + (1+s)\text{Li}_2(-s) + (1-s)\text{Li}_2(s)\Big\} 
     \label{eq:c_eff}
\end{align}
with
\begin{align}
s = \frac{1}{\cosh{4\beta}} = \frac{\left(1-\lambda ^2\right)^2}{\lambda ^4+6 \lambda ^2+1}. 
\end{align}
Here, $\text{Li}_2(z) \equiv \int_z^0 \frac{\log(1-t) dt}{t}$ denotes the dilogarithm function. $\Ceff$ gradually decreases from $c=1/2$ as we increase the measurement strength $\lambda$ from $0$. \cmjadd{Based on the connection between the EE and the twist fields explained above, $\Ceff(\lambda)$ is directly related to scaling dimension of the twist fields in the limit $R\rightarrow 1$, which is a function of the strength of the defect parametrized by $\lambda$.} 

\begin{figure}[tb]
\centering
\captionsetup{justification = RaggedRight}
\includegraphics[width = .9\linewidth]{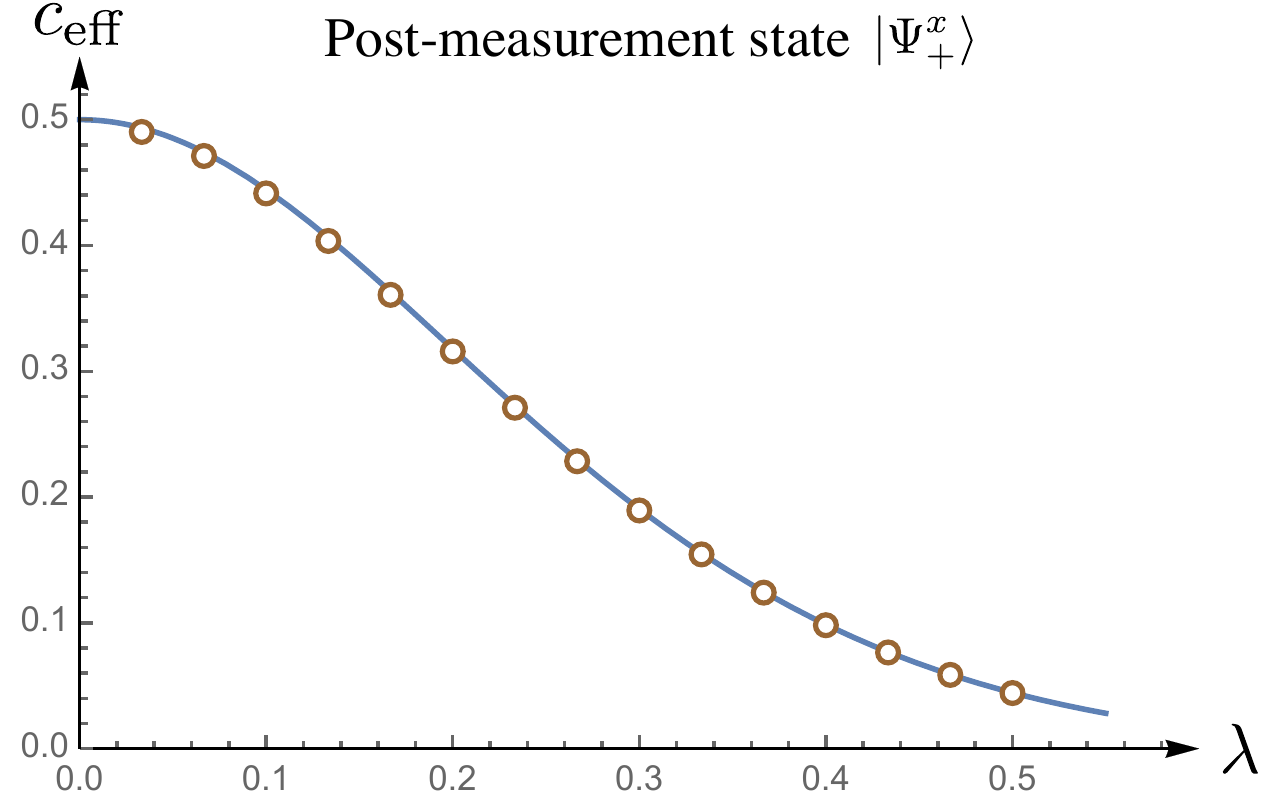}
\caption{Analytic (solid line) and numerical (circles) results for $c_\text{eff}$ of post-measurement state $|\Psi_+^x\rangle$. }\label{fig:ceff1}
\end{figure}

Before we explain the derivation of $\Ceff$, we compare Eqs. \eqref{eq:EE_log_scaling_x_measured} and \eqref{eq:c_eff} with the numerical simulations. For different values of $\beta = {\rm arctanh} \lambda$, the effective central charge associated with the logarithmic scaling of the EE of $|\Psi_+^x\rangle$ can be numerically extracted using the MPS-based method. The details of the numerical procedure based on MPS are provided in App. \ref{app:dmrg}. As shown in Fig. \ref{fig:ceff1}, the analytical expression Eq. \eqref{eq:c_eff} matches perfectly with the effective central charge $\Ceff$ extracted from numerical calculations. There is an alternative numerical method for the study of $|\Psi_+^x\rangle$. We can map the cTFIM of length $L$ to a one-dimensional Majorana-fermion chain with $2L$ sites. We numerically calculate the EE of the post-measurement state $|\Psi_+^x\rangle$ in the Majorana-fermion representation using the covariance matrix formulation. More details of this mapping and the covariance matrix formulation is provided in App. \ref{app:Majorana}. The numerical simulation using the Majorana-fermion representation for $L=512$ also yields the same result as shown in Fig. \ref{fig:ceff1}.

\begin{figure}[tb]
 \captionsetup{justification = RaggedRight}
\includegraphics[width = .9\linewidth]{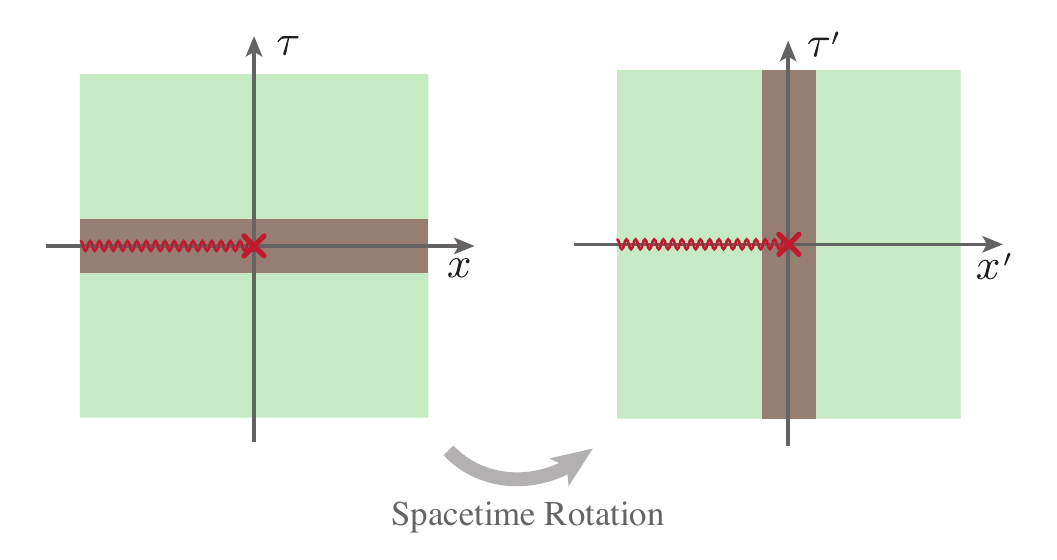}
\caption{ Path integral representation of half-system reduced density matrix $\rho_A$ for post-measurement state $\ket{\Psi^x_+}$ (left panel) and the cTFIM with a bond defect (right panel). $A=(-\infty,0)$ represents half of the infinite spin chain. The two systems (on two panels) are related by a spacetime rotation.}\label{fig:spacetime_rotation}
\end{figure}

Now, we discuss how to obtain the analytical expression Eq. \eqref{eq:c_eff}. In the $R$-replica field-theoretic treatment, the half-system EE in the state $|\Psi^x_+\rangle$ is associated with the one-point function of the twist field $\mathcal{T}_R$ inserted in the middle of the system and at $\tau =0$ (as indicated by the red cross in the left panel of Fig. \ref{fig:spacetime_rotation}) whose associated branch cut follows the constant-time slice at $\tau = 0$. We can perform a $90^\circ$ spacetime rotation of the path integral (see the right panel of Fig. \ref{fig:spacetime_rotation}). The rotated path integral effectively describes a different one-dimensional spin chain with a static Hamiltonian tuned to the critical point of the TFIM everywhere along the chain except at a defect located in the middle point of the chain. Let's call this spin chain the ``Ising chain with a defect". We argue that the half-system EE in the state $|\Psi^x_+\rangle$ shares the same universal behavior as the half-system EE of the Ising chain with a defect. Note that in the latter system, an endpoint of the ``half system" is exactly located at the defect. The twist-field-based field-theoretic description of the half-system EE of the state $|\Psi^x_+\rangle$ can be mapped, under a $90^\circ$ spacetime rotation and a re-arrangement of the branch cut (red wavy line in Fig. \ref{fig:spacetime_rotation}), to the field-theoretic description of the half-system EE of the Ising chain with a defect (see Fig. \ref{fig:spacetime_rotation}). The universal behavior of the half-system EE in both cases is governed by the scaling dimension of the twist field $\mathcal{T}_R$ located within the perturbed region (brown region of Fig. \ref{fig:spacetime_rotation}) of the spacetime. The orientation of the branch cut does not matter. From the perspective of the $R$-replica path integral which is effectively conducted on a $R$-sheeted surface (see Fig. \ref{fig:Perturbed_Ising_Twists}), different choices of the orientation of the branch cut do not affect the geometry of the surface. Hence, the EE in the post-measurement state $|\Psi^x_+\rangle$ shares the same universal behavior as that of the Ising chain with a defect.

A careful treatment of the spacetime rotation can be performed 
by using a two-dimensional classical Ising model on a square lattice with a one-dimensional defect as a proxy to the continuum field theory. The technical details are provided in App. \ref{app:lattice}. Using this proxy two-dimensional classical Ising model, the spacetime rotation maps the problem of the half-system EE of the post-measurement state $|\Psi^x_+\rangle$ to that of an Ising chain with a defect whose Hamiltonian is given by taking the Hamiltonian Eq. \eqref{eq:TFIM} and modifying only the coupling $\sigma^z_j \sigma^z_{j+1}$ for one neighboring pair of sites in the middle of the chain
to 
$t \sigma^z_j \sigma^z_{j+1}$. The parameter $t$ of the Ising chain with a defect is related to the measurement strength $\lambda$ via
\begin{align}
    t = e^{-4\beta}=\left(\frac{1-\lambda}{1+\lambda}\right)^2
    \label{eq:t_as_lambda}.
\end{align}
The half-system EE of such an Ising chain with a defect has been studied in Ref. \onlinecite{eisler2010} and shown to have a logarithmic scaling of the same form as Eq. \eqref{eq:EE_log_scaling_x_measured} with a $t$-dependent effective central charge. The proxy classical Ising model also played an important role in obtaining the EE in the Ising chain with a defect. We remark that the mapping that relates the EE of the post-measurement state $\ket{\Psi^x_+}$ and that in the Ising chain with a defect should be viewed as a mapping between the infrared (IR) physics of the two systems. It should not be interpreted as an exact mapping at the lattice scale.

Using the proxy classical Ising model, we obtain Eq. \eqref{eq:t_as_lambda} and, more importantly, the analytical expression Eq. \eqref{eq:c_eff} of the effective central charge $\Ceff$ for the post-measurement state $|\Psi^x_+\rangle$ which results from the weak measurements along the spin axis $v=x$ on the cTFIM ground state $\IsGS$. By the same method, we can show that the effective central charge $\Ceff$ defined via the half-system EE in the post-measurement state $|\Psi^x_-\rangle$ follows exactly the same expression as Eq. \eqref{eq:c_eff}. However, the fact that $|\Psi^x_-\rangle$ and $|\Psi^x_+\rangle$ share the same effective central charge $\Ceff$ can be explained by the observation that the two post-measurement states are related by a Kramers-Wannier duality (see App. \ref{app:lattice} for details). 
However, this observation does not imply that $|\Psi^x_-\rangle$ and $|\Psi^x_+\rangle$ have completely identical IR properties.
In fact, 
they exhibit different behaviors in the correlation function $\langle \sigma_j^z \sigma_{j'}^z\rangle$ as we explain below.

Again, using the same proxy two-dimensional classical Ising model, one can show that the correlation functions $\langle \Psi_\pm^x| \sigma_j^z \sigma_{j'}^z |\Psi_\pm^x \rangle \sim |j-j'|^{-2\Delta_{z,\pm}}$ in the post-measurement states $|\Psi_\pm^x \rangle$ follow power-law behavior with exponents $\Delta_{z,\pm}$ that vary continuously as we change the measurement strength $\lambda$:
\begin{align}
    \Delta_{z,\pm} = \frac{2}{\pi^2}\arctan^2\left(\left(\frac{1\pm\lambda}{1\mp\lambda}\right)^2 \right).
    \label{eq:DeltaZ_post2}
\end{align}
Note that the expression differs for the two post-measurement states $|\Psi^x_+\rangle$ and $|\Psi^x_-\rangle$.

\begin{figure}[tb]
\centering
 \captionsetup{justification = RaggedRight}
\includegraphics[width = .9\linewidth]{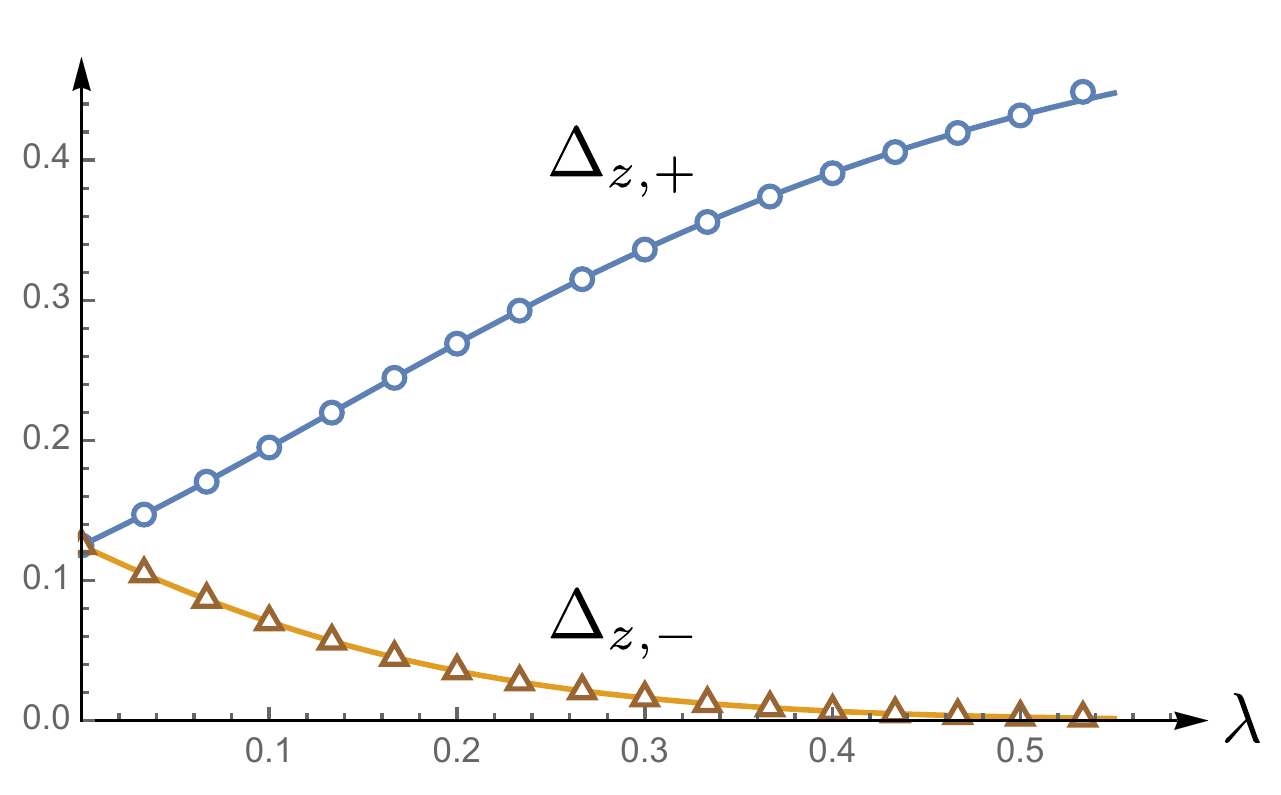}
\caption{$\Delta_{z,\pm}$ extracted from the numerical calculation of $\langle\Psi_\pm^x| \sigma_j^z \sigma_{j'}^z |\Psi_\pm^x \rangle$ are plotted as blues circles and brown triangles for different values of measurement strength $\lambda$. The analytical results in Eq. \eqref{eq:DeltaZ_post2} 
are plotted as solid lines.}\label{fig:DeltaZ_post}
\end{figure}

Numerics confirms this analytical result. In Fig. \ref{fig:DeltaZ_post}, we plot the exponents $\Delta_{z,\pm}$ (blue circles and brown triangles) extracted from the numerically calculated correlation functions $\langle \Psi_\pm^x| \sigma_j^z \sigma_{j'}^z |\Psi_\pm^x \rangle \sim |j-j'|^{-2\Delta_{z,\pm}}$. They exactly match the analytical expression Eq. \eqref{eq:DeltaZ_post2} which is plotted as the blue and brown lines in Fig. \ref{fig:DeltaZ_post}. The numerical simulation is performed on a spin chain of length $L=256$ represented as a Majorana-fermion chain with 512 sites.

\cmjadd{Note that, even though the effective central charge $\Ceff$ and the exponents $ \Delta_{z,\pm} $ vary continuously as $\lambda$ changes, we should view them as the universal properties of the one-dimensional exactly marginal defects in the 1+1d Ising CFT, which are parameterized by only one parameter, i.e. the defect strength. In settings with more general measurement protocols, as long as the measurements perturbed the Ising CFT ground state through the primary field $\phi^x$, the uniform post-measurement states are expected to exhibit the same $\Ceff$ and $ \Delta_{z,\pm} $ upon the correct identification of the defect strength. This universal behavior is analogous to Luttinger liquid, where the low-energy properties are solely determined by the Luttinger parameter, regardless the details of the specific lattice realization.
}

We close this section with some comments on the post-measurement states $|\Psi^v_\pm\rangle$. To prepare the states $|\Psi^v_\pm\rangle$ from measurements, post-selections on the measurement outcomes $\{\m_j\}$ are required. While there is no scalable implementation of the required post-selection in the thermodynamical limit, it is still interesting to investigate if the universal behavior found in this section can be observed in a finite-size near-term quantum device \cite{PreskillNISQ}. We defer this investigation for future study. Instead of being viewed as the results of measurements, $|\Psi^v_\pm\rangle$ can also be thought of as the result of the finite-time evolution with respect to a non-Hermitian Hamiltonian $\sum_j \i\sigma_j^v$, starting from the cTFIM ground state $\IsGS$. The time duration of the evolution $\beta$ is given by the measurement strength via $\beta = {\rm arctanh}(\lambda)$. Hence, the results in this section can also be viewed as the consequences of non-unitary quantum dynamics with the initial state $\IsGS$. At a finite time $\beta$, the acquired state $\ket{\Psi^v_+}$ has an interesting entanglement structure indicated by its effective central charge $\Ceff$. In the long-time limit $\beta\rightarrow \infty$, the state $\ket{\Psi^v_+}$ becomes trivial. We remark that there are also one-dimensional spin chains with different non-unitary dynamics that exhibit interesting entanglement structures in the long-time limit \cite{LuSpacetimeDual,IppolitiSpaceTimeDual,Turkeshi2021IsingNoClicks,Turkeshi2022Reset,Turkeshi2023}.

\section{Born-rule measurements}\label{sec:born}

In this section, we focus on weak Born-rule measurement along the spin-$x$ axis. That means we study the EEs of the post-measurement states $|\Psi^x_{\{\m_j\}}\rangle$ averaged over all possible measurement outcomes $\{\m_j = \pm\}$ with respect to their Born-rule probability Eq. \eqref{eq:Born-Rule_Chain}. 

In the following, we discuss the field theory representation of the Born-rule averaged EE and present our numerical results for effective central charge extracted from the average EE. 

Consider the spin chain of length $L$ which lies along the interval $[-L/2, L/2]$. This chain is divided into two halves $A = (-L/2, 0]$ and $B = (0, L/2)$. For each given measurement outcome $\{\m_j\}$, an un-normalized version of the reduced density matrix on the half-system $A$ is given by
\begin{align}
    \rho_{\{\m_j\},A} &=\Tr_B\left( \prod_j K^x_{j,\m_j}\ket{\Omega} \bra{\Omega} \prod_{j'} K^{x\dag}_{j',\m_{j'}} \right)
    \nonumber \\
    & \propto \Tr_B\left(\ket{\Psi^x_{\{\m_j\}} }\bra{\Psi^x_{\{\m_j\}} }\right).
    \label{eq:rhoA}
\end{align} 
``$\Tr_{A/B}$" represent the trace over the degrees of freedom in the subsystem $A$ and $B$ respectively.
In the following, We will suppress the subscripts of $\rho_{\{\m_j\},A}$ and write it as $\rho$ to simplify the notation without causing confusion. 

The average half-system EE under the Born-rule measurements, denoted as $\mathbb{E}_{\rm b} \,  S(A) $, can be written as
\begin{align}
&\mathbb{E}_{\rm b} \,  S(A) 
= \sum_{\{\m_j=\pm\}} p(\{\m_j\}) S_{|\Psi^x_{\{\m_j\}}\rangle} (A)
\nonumber \\
&= -\sum_{\{\m_j=\pm\}} p(\{\m_j\})\Tr_A \left(\frac{\rho }{\Tr_A(\rho) } \log\left(\frac{\rho}{\Tr_A (\rho)}\right)\right),\label{eq:bornee}
\end{align}
where $\mathbb{E}_{\rm b}$ represents the averaging with respect to the Born-rule probability $p(\{\m_j\})$ given in Eq. \eqref{eq:Born-Rule_Chain} which can be rewritten as
\begin{align}
   p(\{\m_j\}) & =  \Tr_A(\rho )
   \nonumber \\
   & = \Tr \left( \prod_j K^x_{j,\m_j}\ket{\Omega} \bra{\Omega} \prod_{j'} K^{x\dag}_{j',\m_{j'}} \right).
\end{align}
Here, ``$\Tr$" without any subscript indicates the trace over the degrees of freedom in the entire chain.

Now, we discuss the formal field theory representation of this problem. We assume an infinite system size, i.e. $L\rightarrow \infty$, for this discussion. Using the replica trick, we can rewrite Eq. \eqref{eq:bornee} as
\begin{align}
   &\mathbb{E}_{\rm b} \,  S(A) 
   \nonumber \\
   &= \lim_{n\rightarrow 1}\lim_{l\rightarrow 0} \frac{1}{l(1-n)}\sum_{\{\m_j\}}\, \left\{\Tr_A(\rho) \, \Tr_A(\rho^n)^l- \Tr_A(\rho)^{nl+1}\right\} \label{eq:bornee2}
\end{align}
Both $\sum_{\{\m_j\}} \Tr_A(\rho)\Tr_A(\rho^n)^l$ and $\sum_{\{\m_j\}} \Tr_A(\rho)^{nl+1}$ can be expressed in the continuum in terms of an $R$-replica path integral with the replica number $R=nl+1$.

Let us first describe the path integral for $\sum_{\{\m_j\}} \Tr_A(\rho)^{nl+1} 
=  \sum_{\{\m_j\}}  \bra{\Omega} \prod_{j} K^{x\dag}_{j,\m_{j}}  K^x_{j,\m_j}\ket{\Omega}^R
$.
The action of this path integral contains $R$ copies of ${\cal S}_{\rm Ising}$, the action of 1+1d Ising CFT, one for each replica. 
The measurements (and the averaging over different measurement outcomes) induce the coupling $\delta {\cal S}_{\rm M}$ between the replicas:
\begin{align}
   e^{-\delta {\cal S}_{\rm M}} = \sum_{\{\m(x) = \pm \}} \exp\left( 
 \tilde{\beta}\int dx\, \m(x) \sum_{\alpha=1}^R (\phi^x_\alpha(x,\tau) +h ^x)\right),
 \label{eq:deltaS_outcome_summed}
\end{align}
where $\phi^x_\alpha$ is the field $\phi^x$ (namely the $\varepsilon$ field of the 1+1d Ising CFT) in the $\alpha$th replica. \cmjadd{The perturbation $\delta {\cal S}_{\rm M}$ can be viewed as introducing a one-dimensional deffect in the $R$-replica Ising CFT.} Note that in the limit $\lim_{n\rightarrow 1}\lim_{l\rightarrow 0}$ in Eq. \eqref{eq:bornee2}, the number of replicas has a limit $R\rightarrow 1$.

The coupling constant $h_x$, independent of the measurement strength $\lambda$, is introduced to account for an important difference between the operator $\sigma_j^x$ of the spin chain and the field $\phi^x = \varepsilon$ of the Ising CFT.  The former has a finite expectation value $\langle \Omega| \sigma^x_j |\Omega\rangle = 2/\pi$ on the cTFIM ground state while the expectation value of the field $\phi^x = \varepsilon$ vanishes on the ground state \footnote{The expectation value $\langle \Omega| \sigma^x_j |\Omega\rangle = 2/\pi$ can be straightforwardly calculated using the Majorana-fermion representation of cTFIM.}. Therefore, in a field theory representation, we should substitute $\sigma^x_j$ with $\phi^x+h^x$ with $h^x>0$. This finite coupling $h^x$ results in an asymmetry between the measurement outcomes $\m(x) = \pm 1$.

We remark that a finite $h^x$ does not affect our previous field-theoretic analysis for the post-measurement states $|\Psi_\pm^x\rangle$. That is because the measurement outcomes $\m_j$ are fixed for $|\Psi_\pm^x\rangle$, and the effect of a finite $h_x$ is removed once we normalize these post-measurement states.

The path integral description of $\sum_{\{\m_j\}} \Tr_A(\rho)\Tr_A(\rho^n)^l$ appearing in Eq. \eqref{eq:bornee2} is given by almost the same $R$-replica field theory as that of $\sum_{\{\m_j\}} \Tr_A(\rho)^{nl+1}$ except that a branch cut needs to be introduced at $\tau = 0$ along the line $(-\infty, 0)$ on the spatial axis (where the subsystem $A$ is located). Crossing the branch cut implements a permutation $T$ of the replica index: $\alpha \rightarrow T(\alpha)$. $T$ is an element of the permutation group $S_R$ whose cycle representation is given by $\big(1,2,...,n\big)\big(n+1,n+2,...,2n\big)...\big((l-1)n+1,(l-1)n+2,...,ln\big)$ where the last cycle is of length one and hence suppressed in the notation. We can view the branch cut as the consequence of the insertion of a twist field $\mathcal{T}_R$ at $(x=0,\tau=0)$.

From the path integral, one can readily argue that the coupling $e^{-\delta {\cal S}_{\rm M}}$ in Eq. \eqref{eq:deltaS_outcome_summed} does not contain any relevant perturbation to the replicated Ising CFT. In the replica limit $R\rightarrow 1$, we can show analytically that even the marginal perturbation is absent, rendering the measurement-induced perturbation $e^{-\delta {\cal S}_{\rm M}}$ irrelevant under RG. The details of this RG analysis are provided in App. \ref{app:ceff}. A direct consequence is that, for Born-rule measurements, the effective central charge should remain at $\Ceff = \frac{1}{2}$ regardless of the measurement strength. As shown below, this conclusion is verified by our MPS-based numerical simulations.

Using the MPS method, we calculate the average half-system EE for an infinite spin chain. The bond dimension of the MPS serves as a cut-off for the EE. The effective central charge can be extracted from the dependence of the EE on the bond dimension. Ideally, the
measurement outcome (and the associated Kraus operator) on each site of the infinite chain should be independent. In the numerical simulation, we cannot perform independent measurements on an infinite number of sites. Instead, we take an approximation by assuming a periodicity in the measurement outcomes with a unit cell $L$, namely $\m_j= \m_{j+L}$. The ideal case can then be approached by taking $L \rightarrow \infty$. In the following simulations, we take $L=120$.

Instead of measuring all the spins at once, we measure the spins one by one. For each spin, we follow the Born-rule probability Eq. \eqref{eq:Born-ruleX} to choose a measurement outcome, obtain the resulting post-measurement state following Eq. \eqref{eq:wavefunction_collapse}, and then move on to the measurement of the next spin. One can show straightforwardly that the measurement outcomes $\{\m_j\}$ generated in this one-by-one measurement procedure follow exactly the (joint) Born-rule probability Eq. \eqref{eq:Born-Rule_Chain}. Once all the spins are measured starting from the cTFIM ground state $\IsGS$, we obtain a post-measurement state $|\Psi_{\{\m_j\}}\rangle$ whose half-system EE can be directly numerically calculated.

To obtain the average half-system EE, one can, in principle, repeat the one-by-one measurement procedure starting from $\IsGS$ over and over. To speed up the convergence, we use the following trick. When we finish one round of measurements of all the spins starting from $\IsGS$, we obtain a specific set of measurement outcomes $\{\m_j\}$. We can simultaneously calculate the half-system EE for post-measurement states $|\Psi_{\{\m'_j\}}\rangle$ with $\{\m'_j= \m_{j+k}\}$ which are obtained from $\{\m_j\}$ by shifting the site index by  $k=0,1,..,L-1$. We then average the half-system EE of all post-measurement states $|\Psi_{\{\m'_j\}}\rangle$ with different $k$'s. That is, we average half-system EE
over $L$ different measurement outcomes $\{|\Psi_{\{\m_j\}}\rangle, |\Psi_{\{\m_{j+1}\}}\rangle\},...,|\Psi_{\{\m_{j+L-1}\}}\rangle\}$. In other words, we trade the statistical average over different $\{\m_j\}$ by a spatial average. Without changing $L$, the MPS method allows us to extract the effective central charge $\Ceff$ from the dependence of the EE on the correlation length, both of which are cut off by the MPS bond dimension. By varying the bond dimension, their dependence can be extracted. More details can be found in App. \ref{app:dmrg}. Conceptually, our approach for the average EE is similar to calculating the half-system EE for a finite spin chain, whose length is given by the correlation length of the MPS representation.

\begin{figure}
    \captionsetup{justification = RaggedRight}
    \includegraphics[width=0.4\textwidth]{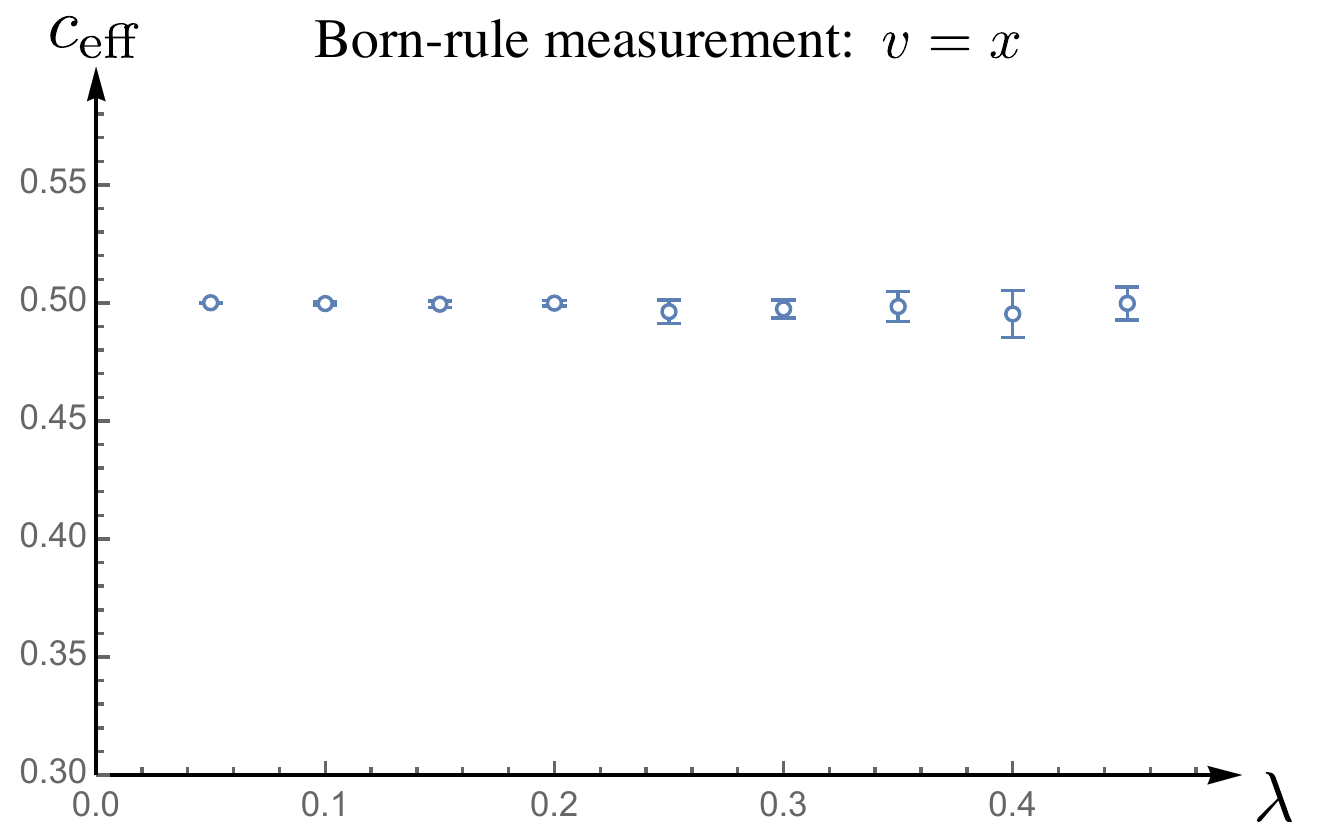}
    \caption{For the Born-rule measurement along the spin-$x$-axis, the average EE follows a logarithmic scaling. The numerically extracted effective central charge is plotted as a function of $\lambda$.
    }
    \label{fig:born}
\end{figure}

In Fig. \ref{fig:born}, we present the extracted $\Ceff$ for various measurement strengths $\lambda$. The error bars represent the fluctuation of the extracted $\Ceff$ from 10 different sets of $\{\m_j\}$ (not related by shifting) with each set generated by the one-by-one measurement procedure starting from $\IsGS$. We find that, with the Born-rule measurement along the spin-$x$-axis, the effective central charge $\Ceff$ appears to be independent of the measurement strength $\lambda$ and takes the value $0.5$, which is in strong in contrast to the post-measurement states $|\Psi_{\pm}^x\rangle$. The fluctuations between different sets of $\{\m_j\}$ are very small, indicating that the spatial average on the infinite spin chain 
with a $L=120$ unit cell for the measurement outcomes provides a good approximation to the behavior of averaged post-measurement EE in the true infinite system limit (with $L\rightarrow\infty$).

The alternative numerical method using the Majorana-fermion representation of a length-512 spin chain produces consistent results as those shown in Fig. \ref{fig:born}. In this method, we calculate the average EE for different interval sizes $l$ and extract $\Ceff$ by fitting the average EE to $\log l$.

The independence of $\Ceff$ on $\lambda$ is somewhat surprising given that the usual (and possibly naive) expectation is such that measurements of non-overlapping (and, hence, commuting) observables generally reduce the entanglement in the system. This expectation is correct for the post-measurement states $\ket{\Psi^v_{\pm}}$ but at odds with our findings for the average EE with Born-rule measurements. In terms of the field theory, the independence of $\Ceff$ on the measurement strength $\lambda$ substantiates our analytical result (detailed in App. \ref{app:ceff}) that the coupling $\delta {\cal S}_{\rm M}$ in Eq. \eqref{eq:deltaS_outcome_summed} is irrelevant in the replica limit $R\rightarrow 1$ and, hence,  does not affect the scaling behavior of the twist fields $\mathcal{T}_R$ in this replica limit.

The Born-rule probability $p(\{\m_j\})$ contains non-trivial correlations between the measurement outcome $\m_j$ at different sites $j$. In the following section, we show that one can approximate the Born-rule probability $p(\{\m_j\})$ by an un-correlated probability distribution $p'(\{\m_j\})$. Under this un-correlated probability distribution for the post-measurement-state ensemble, the feature that $\Ceff$ is independent of the measurement strength $\lambda$ is retained.

\section{Forced measurements}\label{sec:force}
In this section, we consider the average behavior of the EE after the forced measurement on the cTFIM ground state $\IsGS$. That is, we average the EE over all possible measurement outcomes $\{\m_j = \pm\}$ with equal probability. For this section, the weak measurement on each spin is still along the spin-$x$ axis. We show that under the forced measurement, the average half-system EE still exhibits the logarithmic scaling $\sim \frac{\Ceff}{6} \log (L/2)$ where the effective central charge $\Ceff$ depends non-trivially on the measurement strength. Moreover, we introduce a biased version of the forced measurement where an uncorrelated bias towards one of the two measurement outcomes $\m_j=\pm$ on every site is introduced. With a proper choice of the bias, we can recover a feature of the Born-rule averaged EE, that is the independence of $\Ceff$ on the measurement strength $\lambda$. We show that the probability distribution of measurement outcomes $\{m_j\}$ with this biased forced measurement serves as a mean-field approximation to the Born-rule probability distribution as far as the average EE goes.

For the chain of length $L$ divided into two halves $A=(-L/2,0]$ and $B = (0,L/2) $, the average EE $\mathbb{E}_{\rm f} \,  S(A) $ of the half system $A$ under the forced measurement can be expressed as 
\begin{align}
&\mathbb{E}_{\rm f} \,  S(A) 
\nonumber \\
& = - \frac{1}{2^L}\sum_{\{\m_j=\pm\}}\Tr_A \left(\frac{\rho }{\Tr_A(\rho )} \log\left(\frac{\rho}{\Tr_A (\rho)}\right)\right),\label{eq:forceee1}
\end{align}
where $\rho$ follows the definition in Eq. \eqref{eq:rhoA} and $\mathbb{E}_{\rm f}$ represents the average over all measurement outcomes with equal probability.

Again, by applying the replica trick, we can write 
\begin{align}
&\mathbb{E}_{\rm f} \,  S(A) 
\nonumber \\
& = \frac{1}{2^L} \lim_{n\rightarrow 1}\lim_{l\rightarrow 0} \frac{1}{l(1-n)}\sum_{\{\m_j\}}\, \left\{\Tr_A(\rho^n)^l- \Tr_A(\rho)^{nl}\right\} \label{eq:forceee2}
\end{align}
Both $\sum_{\{\m_j\}}\Tr_A(\rho^n)^l$ and $\sum_{\{\m_j\}}\Tr_A(\rho)^{nl}$ can be formally expressed in terms of an $R$-replica path integral where the replica number is given by $R = n l$. In the limit $\lim_{n\rightarrow 1}\lim_{l\rightarrow 0}$, the number of replicas $R\rightarrow 0 $ (as opposed to $R\rightarrow 1 $ in the case of Born-rule measurement). The path integral formulation of $\sum_{\{\m_j\}}\Tr_A(\rho)^{nl}$ is identical to the $R$-replica path integral formulated in Sec. \ref{sec:born} except that $R=nl$. Hence, the path integral action is given by $R=nl$ copies of 1+1d Ising CFT coupled to each other via Eq. \eqref{eq:deltaS_outcome_summed}. For $\sum_{\{\m_j\}}\Tr_A(\rho^n)^l$, a branch cut at $\tau =0$ along the interval $(-\infty, 0]$ is needed (assuming the system size $L$ is infinite). Crossing the branch cut results in a permutation $T'$ of the replica index. $\alpha \rightarrow T'(\alpha)$. $T'$ is an element of the permutation group $S_R$ whose cycle representation is given by $\big(1,2,...,n\big)\big(n+1,n+2,...,2n\big)...\big((l-1)n+1,(l-1)n+2,...,ln\big)$ (without any additional trivial cycles).

\begin{figure}
    \captionsetup{justification = RaggedRight}
    \includegraphics[width=0.4\textwidth]{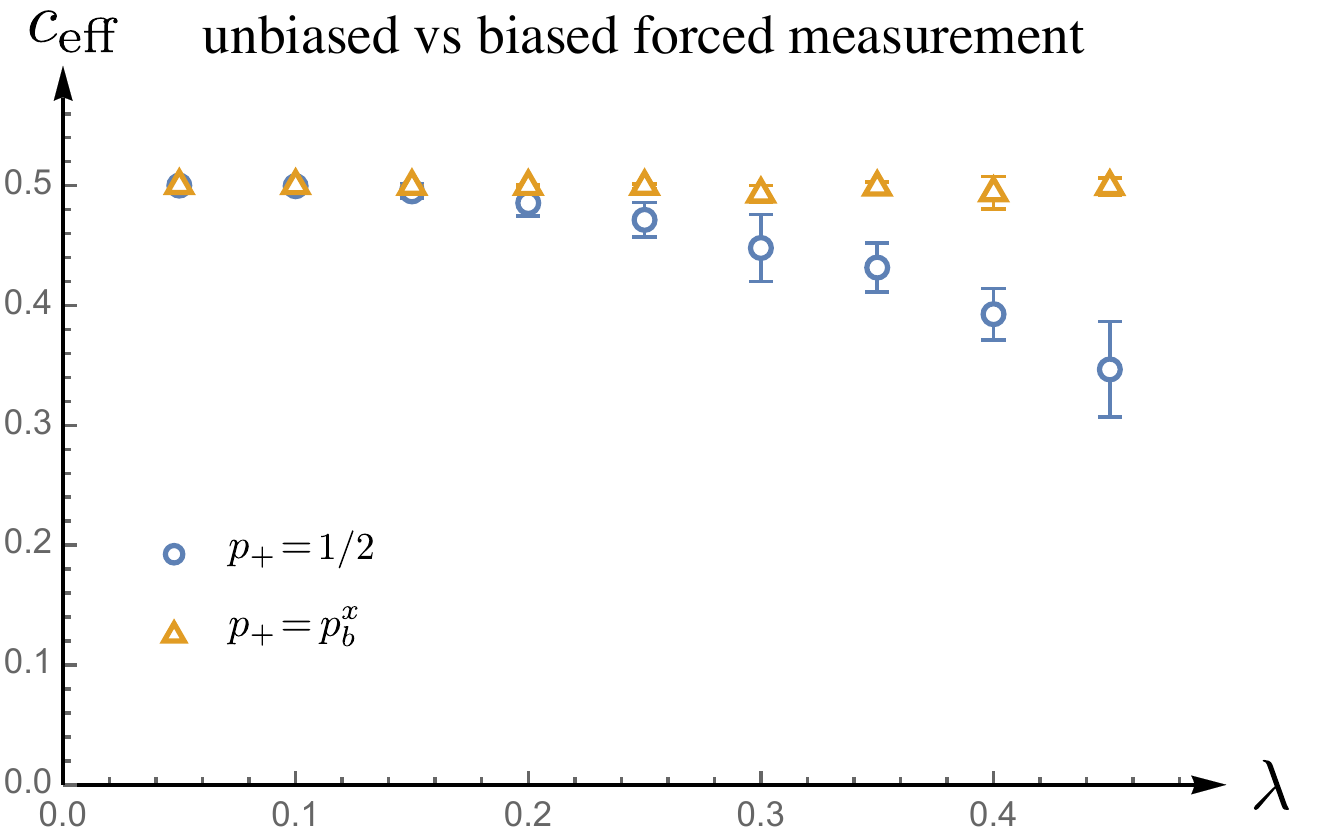}
    \caption{
    The numerically extracted effective central charges $\Ceff$
    from the forced measurement ($p_+ = 1/2$) and from the biased forced measurement with $p_+ = p^x_b$
    are plotted as functions of $\lambda$.
    } 
    \label{fig:forced}
\end{figure}

The numerical calculation of the average EE is similar to the case with Born-rule measurements. This numerical calculation is also performed using the MPS-based method on an infinite spin chain with periodic measurement outcomes with a unit cell $L=120$. The measurement outcomes $\{\m_j\}$ are also sampled using the one-by-one measurement procedure. Unlike the Born-rule case, for the forced measurement of every spin, we sample the measurement outcome $\m_j=+$ with probability $p_+ = 1/2$ and $\m_j=-$ with probability $p_- = 1-p_+ = 1/2$. Once all the spins are measured once starting from the cTFIM ground state $\IsGS$, we can calculate the average half-system EE. It is found to follow the logarithmic scaling still. We extract the effective central charge $\Ceff$ in the same way as we did in the Born-rule case. $\Ceff$ for various measurement strength $\lambda$ is plotted as the blue circles in Fig. \ref{fig:forced}. The error bars represent the fluctuation of the extracted $\Ceff$ from 10 different sets of measurement outcomes $\{\m_j\}$ with each set independently generated using the one-by-one measurement procedure starting from $\IsGS$. We find that $\Ceff$ has a non-trivial dependence on the measurement strength $\lambda$. 

The analytical approach to study the dependence of $\Ceff$ on the forced measurement strength $\lambda$ is explained in detail in App. \ref{app:ceff}. The analytical approach begins with the observation that the perturbation induced by the forced measurements also takes the form Eq. \eqref{eq:deltaS_outcome_summed}, and it does not contain any RG-relevant terms. The next step is to extract the exactly marginal terms contained in the perturbation in the replica limit $R\rightarrow 0$. The effective central charge depends on the exactly marginal terms in a similar way as it does in the case of the post-measurement states $|\Psi^x_\pm\rangle$ discussed in Sec. \ref{sec:post-selection_EE}. As shown in App. \ref{app:ceff}, this analytical approach provides an analytical expression of the effective central charge that matches the numerical simulations.

We can generalize the one-by-one forced measurement procedure to a biased version with a general value of $p_+$ and $p_-=1-p_+$. When $p_+$ deviates from $1/2$, the forced measurement imposes a bias towards one of the measurement outcomes $\m = \pm$. For a general $p_+$, we have a biased forced measurement where different sets of measurement outcomes $\{\m_j\}$ are sampled based on the probability distribution
\begin{align}
    p'(\{\m_j\}) = p_+^{N_+} (1-p_+)^{N_-},
    \label{eq:biased_forced_prob}
\end{align}
where $N_\pm$ is the number of sites where the measurement outcome $\m_j$ is $\pm$. For this probability distribution $p'(\{\m_j\})$, there is no correlation between the measurement outcomes at different sites. We can use $p'(\{\m_j\})$ as a ``mean-field" approximation to the Born-rule probability $p(\{\m_j\})$ in Eq. \eqref{eq:Born-Rule_Chain}. By minimizing the Kullback–Leibler divergence \cite{KL1951} between the two probability distributions $p$ and $p'$, we find that the optimal mean-field approximation is achieved at 
\begin{align}
    p_+ = p^x_b \equiv \frac{1}{2}+ \frac{2\lambda}{\pi (1+\lambda^2)}.
    \label{eq:optimal_p+}
\end{align}
Note that $p^x_b$ is exactly the probability of finding the measurement outcome $\m=+$ when we perform only a single-site measurement along the spin-$x$-axis on the cTFIM ground state $\IsGS$. The expression of $p_b^x$ can be obtained from Eq. \eqref{eq:Born-ruleX} and the fact that $\langle \Omega | \sigma^x_j \IsGS = 2/\pi$. More details of the Kullback–Leibler divergence and its minimization are provided in App. \ref{app:OptimalBias}. Note that with $p_+=p_b^x$, the probability distribution $p'(\{m_j\})$ and the Born-rule counterpart $p(\{m_j\})$ yield the same distribution when we reduce them to a single site (by summing over all possible values of $\m_j=\pm$ on other sites). Hence, $p'$ provides a mean-field approximation to $p$. For the biased forced measurement with $p_+ = p_b^x$, the numerically extracted effective central charge is plotted in Fig. \ref{fig:forced} as the brown triangles and is found to be independent of the measurement strength $\lambda$, which is a feature shared by the case of the Born-rule measurement. The numerical method to extract the effective central charge for the case with biased forced measurements parallels that of the forced measurements. 

\begin{figure}
    \captionsetup{justification = RaggedRight}
    \includegraphics[width=0.4\textwidth]{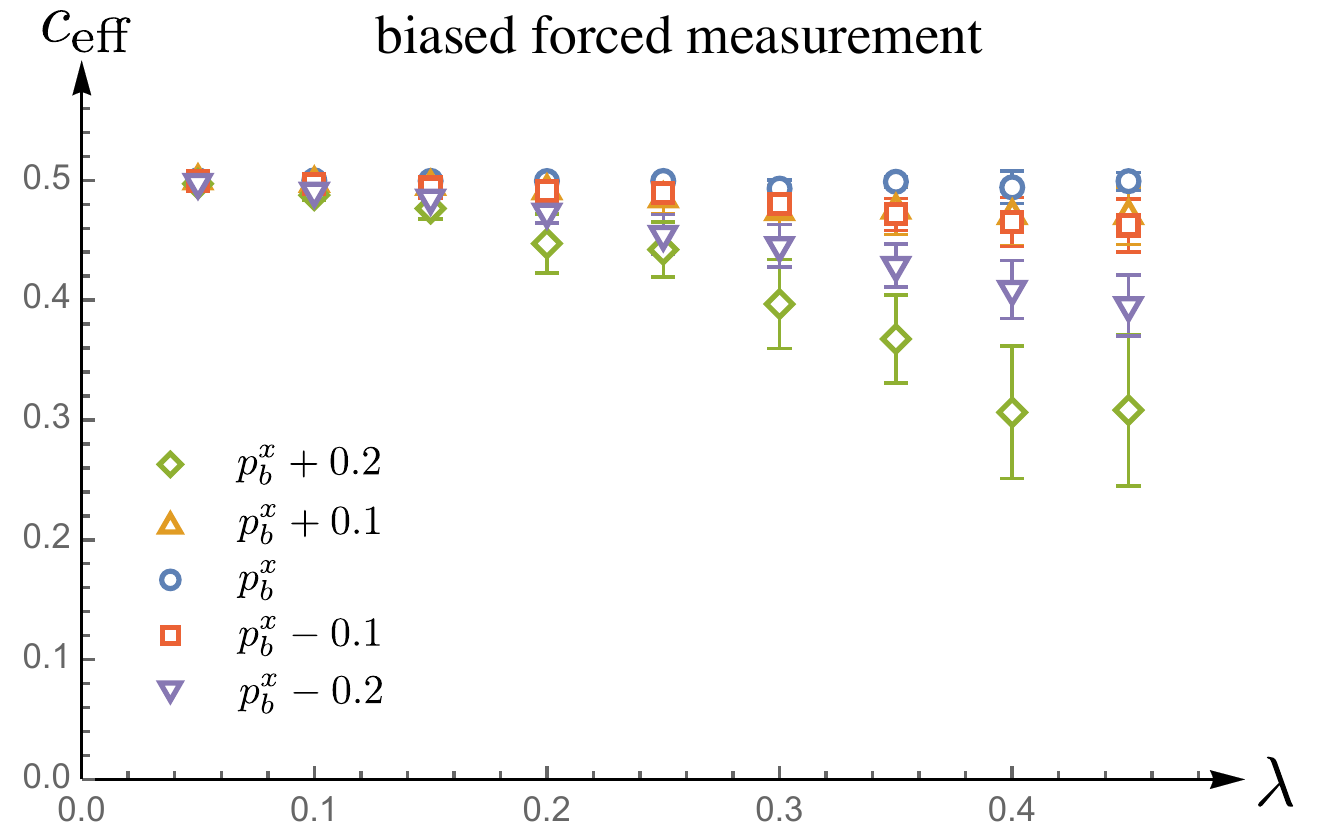}
    \caption{The numerically extracted effective central charges $\Ceff$ from the biased forced measurement with  $p_+ = p^x_b, p^x_b\pm 0.1, p^x_b \pm 0.2$
    are plotted as functions of $\lambda$.} 
    \label{fig:biased_forced}
\end{figure}

To further demonstrate that $p_+=p_b^x$ produces the optimal mean-field approximation of the Born-rule probability $p(\{\m_j\})$ at the level of EE, we plot the numerically extracted effective central charge for various choices of $p_+ = p_b^x \pm \delta p$ with $\delta p = 0, \pm 0.1, \pm 0.2$ in Fig. \ref{fig:biased_forced}. Any deviation from $p_+ = p_b^x$ results in some non-trivial dependence on the measurement strength of the effective central charge $\Ceff$. 

The fact that the bias forced measurement $p'(\{\m_j\})$ with $p_+=p_b^x$ procedures the same feature, namely the independence of $\Ceff$ on $\lambda$, as the Born-rule measurement suggests that the non-trivial correlation contained in the Born-rule probability $p(\{\m_j\})$ between the measurement outcomes $\m_j$'s on different sites is unimportant for the effective central charge $\Ceff$. The numerical results in Fig. \ref{fig:biased_forced} show that the field theory that described the cTFIM ground state $\ket{\Omega}$ under biased forced measurements 
contains a non-trivial marginal perturbation (to the replicated Ising CFT) for a generic value $p_+$ in the replica limit $R\rightarrow 0$. \cmjadd{In other words, a marginal one-dimensional defect with finite strength is present for generic $p_+$. The defect strength is turned off}
when $p_+=p_b^x$. Our analytical study of the measurement-induced perturbation (detailed in App. \ref{app:ceff}) can be generalized to the case with biased forced measurements. The dependence of the effective central charge on both $p_+$ and the measurement strength $\lambda$ are calculated in App. \ref{app:ceff} and are shown to agree with our numerical results.

\section{Summary and Outlook}
\label{sec:summary}

In this paper, we study the effect of weak measurements on the universal properties of the ground state $\IsGS$ of the one-dimensional cTFIM with a focus on the behavior of the post-measurement EE in the system.
Starting from the cTFIM ground state $\IsGS$, we consider the situation where weak measurements are performed on every spin in the system along their spin axis $v=x,z$. For each set of measurement outcomes $\{\m_j\}$, there is a post-measurement state $|\Psi^v_{\{\m_j\}}\rangle$ in the corresponding quantum trajectory. Based on how the ensemble of post-measurement states is sampled, our work contains 3 parts: (1) post-measurement states with spatially uniform measurement outcomes, (2) Born-rule measurement, and (3) forced measurements. \cmjadd{In each part, the post-measurement properties are related to certain one-dimensional defects in the 1+1d Ising CFT or its multi-replica counterpart}.

In the first part, we identify the interesting post-measurement states $|\Psi^v_\pm\rangle$ associated with the uniform measurement outcomes $\{\m_j= +\}$ and $\{\m_j= -\}$ respectively. We show that the properties of $|\Psi^v_\pm\rangle$ can be studied using a 1+1d Ising CFT with a measurement-induced perturbation along a one-dimensional defect. When the measurement is performed with respect to the spin axis $v=x$, the perturbation is exactly marginal. Consequently, the effective central charge $\Ceff$ of $|\Psi^x_\pm\rangle$, defined via the logarithmic scaling of the EE, can be continuously deformed as a function of the measurement strength $\lambda$. We obtain the analytical expression Eq. \eqref{eq:c_eff} of $\Ceff$ using a combination of field-theoretic analysis and a mapping between the post-measurement state $|\Psi^x_\pm\rangle$ and the ground state of an Ising chain with a defect studied in Ref. \onlinecite{eisler2010}. Moreover, we show that the correlation function $\langle \sigma_j^z \sigma_{j'}^z\rangle \sim |j-j'|^{-2\Delta_{z,\pm}} $ in the post-measurement states $|\Psi^x_\pm\rangle$ exhibit continuously tunable exponents $\Delta_{z,\pm}$ whose exact expressions are given in Eq. \eqref{eq:DeltaZ_post2}. \cmjadd{The effective central charge $\Ceff$ and the exponents $\Delta_{z,\pm}$ in the uniform post-measurement states are directly related to universal properties of the one-parameter family of exactly marginal defects in the Ising CFT.} In contrast, when the measurements are conducted with respect to the spin axis $v=z$, we use an RG argument to show that the effective central charge $\Ceff$ immediately vanishes when the measurement strength $\lambda$ is non-zero. All the analytical results of the effective central charge $\Ceff$ and the exponents $\Delta_{z,\pm}$ are compared with those extracted from numerical simulations. 
There are two independent methods for this numerical simulation, one based on MPS and the other based on the Majorana-fermion representation of the spin chain. They both yield consistent results exactly matching the analytical expressions of $\Ceff$ and $\Delta_{z,\pm}$ in Eqs. \eqref{eq:c_eff} and \eqref{eq:DeltaZ_post2}. 

We comment on the post-measurement states $|\Psi^v_\pm\rangle$. In addition to thinking of them as the consequence of measurements followed by post-selecting the measurement outcomes, they can also be thought of as being generated from the cTFIM ground state $\IsGS$ via an evolution with respect to a non-Hermitian Hamiltonian $\sum_j \i\sigma_j^v$ for time $\lambda$. Hence, our results in this part can also be viewed as the consequence of non-unitary quantum dynamics with the cTFIM ground state $\IsGS$ as the initial state.

In the second part, we study the Born-rule measurement where we sample different measurement outcomes $\{\m_j\}$ and their corresponding post-measurement states $|\Psi^x_{\{\m_j\}} \rangle$ following the naturally occurring Born-rule probability $p(\{\m_j\})$ given in Eq. \eqref{eq:Born-Rule_Chain}. For this part, we focus on the measurements with respect to the spin-$x$ axis. We show that the Born-rule averaged EE can be formulated as a problem of a 1+1d $R$-replica Ising CFT coupled to each other by a measurement-induced perturbation along a one-dimensional defect at the limit $R\rightarrow1$. Using numerical simulations based on MPS and the alternative method based on the Majorana-fermion representation of the spin chain, we find that the EE averaged over different post-measurement states $|\Psi^x_{\{\m_j\}} \rangle$ with respect to the Born-rule probability still follows the logarithmic scaling. Our numerical result shows that effective central charge $\Ceff$ is independent of the measurement strength $\lambda$. This behavior can be explained using our analytical study of the measurement-induced defect in the replica limit $R\rightarrow 1$. 
This discovery is surprising because one usually expects the mutually commuting measurements on non-overlapping observables, such as the measurements considered in this work, to reduce the entanglement entropy of the system. This result also suggests that post-measurement states $|\Psi^x_\pm\rangle$ are atypical states, as far as the EE scaling goes, among the ensemble of post-measurement states $|\Psi^x_{\{\m_j\}} \rangle$ weighted by their corresponding Born-rule probability $p(\{\m_j\})$. 

Lastly, we study the case of forced measurements (with respect to the spin $x$ axis) where we sample different measurement outcomes $\{\m_j\}$ with equal probability and, more generally, the case of biased force measurements where $\{\m_j\}$ are sampled according to a pre-determined probability distribution $p'(\{\m_j\})$ given in Eq. \eqref{eq:biased_forced_prob} with a tunable bias $p_+$. Like the case with Born-rule measurements, the average EE with forced measurements can also be formulated as a 1+1d $R$-replica Ising CFT coupled to each other by a measurement-induced perturbation along a one-dimensional defect. The difference is that the replica limit is $R\rightarrow 0$ (as opposed to $R\rightarrow 1$ in the Born-rule case). Based on our numerical simulations, the average EE with forced measurements (which is equivalent to choosing $p_+=1/2$) retains its logarithmic scaling but exhibits an effective central charge $\Ceff$ that decreases continuously as the measurement strength $\lambda$ grows. In the case of biased force measurements with a generic choice of $p_+$, a similar dependence of $\Ceff$ on $\lambda$ is found. Interestingly, for the $p_+$ value given in Eq. \eqref{eq:optimal_p+}, the biased-forced-measurement probability distribution $p'(\{\m_j\})$ for the measurement outcomes $\{\m_j\}$ provides a mean-field approximation of the Born-rule probability distribution $p(\{\m_j\})$. In this case, the average EE with respect to the probability distribution $p'(\{\m_j\})$ exhibits an effective central charge $\Ceff$ independent of the measurement strength $\lambda$, a feature shared by the average EE with Born-rule measurements. The dependence of the effective central charge on the bias $p_+$ and the forced measurement strength $\lambda$ can be analytically calculated by studying the associated measurement-induced defect in the corresponding replica limit.

\dmadd{Our investigation into the measurement of cTFIM not only provides valuable insights but also paves the way for several intriguing avenues of future research. To highlight a few of these promising directions:
1. Expanding our approach to encompass other 1+1d CFTs and critical states in higher dimensions or those following volume-law EE.
2. Exploring the possibility of measurement-induced transitions in other entanglement-related quantities.
3. Gaining a deeper understanding of the efficiency and effectiveness of the measurement process itself by considering other metrics besides EE.
4. Extending our study to address quantum dynamics scenarios that involve continuous monitoring and measurement.
}

 \cmjadd{In the microscopic model, we can consider a general measurement that is defined by a Kraus-operator set satisfying the POVM condition.} From a continuum perspective, as mentioned earlier, the effect of measurement on a one-dimensional critical ground state can be captured by a 1+1d CFT or its multi-replica counterpart with a measurement-induced perturbation on a one-dimensional defect. \cmjadd{Different measurements can lead to different types of defects.} An interesting case is given by measuring observables with scaling dimension 1 in the CFT. An example is the current operator in a one-dimensional critical system with a conserved charge. When considering the post-measurement states with spatially uniform measurement outcomes, the measurement-induced perturbation to the 1+1d CFT is marginal in the RG sense. A possible consequence of such measurement is that the effective central charge associated with the logarithmic scaling of post-measurement EE changes continuously as a function of the measurement strength, just like the cases studied in Sec. \ref{sec:post-selection_EE}. \cmjadd{For analyzing the average properties in the post-measurement states, in particular the effective central charge, our work has demonstrated the effectiveness of the method that combines field-theoretic and replica-trick-based techniques. It is worthwhile to generalize this method to understand the behavior of the average post-measurement EE for more general critical systems.}

In a 1+1d CFT, one can extract the central charge, which is an intrinsic quantity associated with the CFT, from the logarithmic scaling of the EE in the ground state. In a 2+1d critical state with an emergent Lorentz symmetry, the EE on a circular region of radius $L$ follows the universal scaling $S(L) \sim \alpha L - \gamma $ where subleading term $\gamma$ is an intrinsic quantity to the 2+1d CFT that governs the low-energy physics of the system \cite{MyersSinha2010,MyersSinha2011,CasiniHuerta2007,CasiniHuerta2012}. A natural generalization of the study presented in this paper is to understand if the same $S(L) \sim \alpha L - \gamma $ scaling holds for the (average) EE in the post-measurement states obtained from measuring 2+1d critical ground states. If yes, it is interesting to study how $\gamma$ depends on the measurement strength. \cmjadd{More generally, critical ground states are less entangled compared to highly-excited states, which generally exhibit volume-law entanglement entropy scaling. It would be interesting to find cases where the effect of measurements on the highly excited states can be captured by continuum field theories. }


In our study of the cTFIM, the effective central charge $\Ceff$ is either a continuous function of the measurement strength $\lambda$ or vanishes immediately when the measurement strength is non-zero.
As exemplified by recent studies \cite{LeeJianXu2023,BaoFan2023,FanBao2023}, entanglement-related quantities can experience phase transitions induced by finite-strength measurements and related decoherence processes acting on higher-dimensional correlated ground states (at criticality or with topological orders). It is interesting to search for a one-dimensional critical system where the average EE and other entanglement-related quantities experience similar phase transitions at a finite measurement strength.

{\it Acknowledgements} — DM thanks Daniel E. Parker for helpful discussions on numerical simulation. CMJ thanks Bela Bauer, Anna Keselman, and Andreas W. W. Ludwig for the collaboration on an earlier project that partly motivated this study. CMJ thanks Jong Yeon Lee and Cenke Xu for the collaboration on a related project \cite{LeeJianXu2023} that investigates the effect of decoherence and weak measurements on higher-dimensional critical ground states. We thank Eun-Ah Kim for her assistance with computation resources.  
DM is supported by a Bethe/KIC postdoctoral fellowship at Cornell University. This work is supported by a faculty startup grant at Cornell University.

{\it Note added}: Upon completion of this manuscript, we became aware of an independent and related work \cite{ehud2023},  which will appear on arXiv on the same day. 
We thank the authors for informing us of their work in advance.

While completing this manuscript, we also noticed another related independent work \cite{ZouSanHsieh2023}.

\appendix
\section{Details of the MPS simulation}
\label{app:dmrg}
\begin{figure}[h]
    \centering
    \includegraphics[width = 0.45\textwidth]{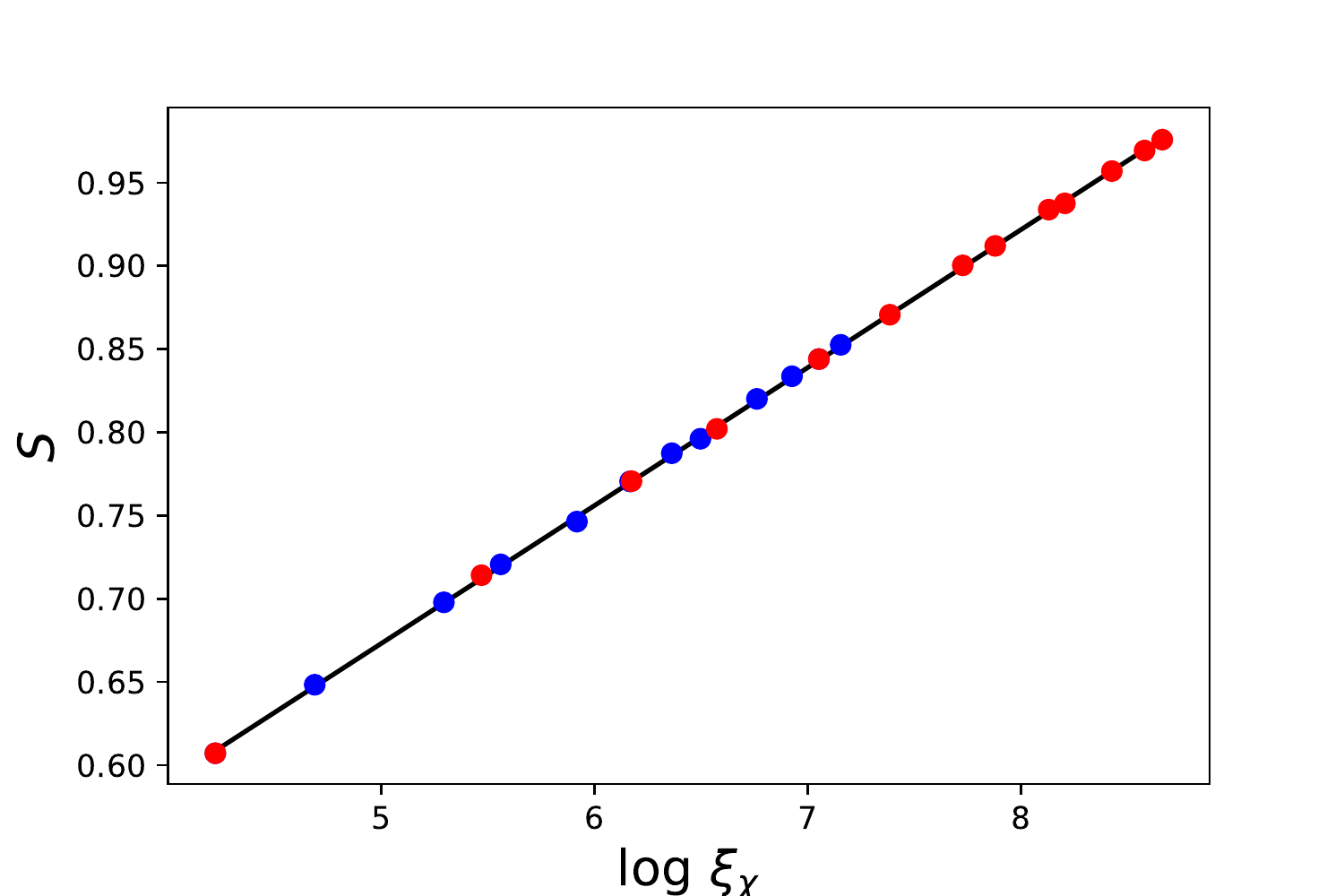}
    \caption{Fitting central charge of TFIM. Blue dots correspond to the bond dimension used in the main text, where $\chi_{max} = 29$. Red dots correspond to another set of bond dimensions, where the maximum is 67. They both give a fit with central charge $c = 0.50$ (black line).}
    \label{fig:c_diff_bond_dim}
\end{figure}

\dmadd{ We first prepare the matrix product state representation of the cTFIM ground state on an infinite chain with different bond dimensions $\chi$ using iDMRG algorithm provided by the TeNPy package \cite{tenpy}. We use the method in Ref. \onlinecite{Pollmann09} for extracting the central charge. Namely, for each $\chi$, we obtain the correlation length  $\xi$ from the second largest eigenvalue of the transfer matrix and the half-chain entanglement entropy $S$ from the Schmidt decomposition. The central charge $c$ can be obtained by fitting $S = \frac{c}{6} \log \xi$ \cite{calabrese2009}.  The maximum bond dimension we use is $ \chi_{max} = 29$, which is similar to the maximum bond dimension used in Ref. \onlinecite{Pollmann09} for cTFIM. We also check the fitting of central charge with larger bond dimensions and the fits converge. (See Fig. \ref{fig:c_diff_bond_dim} for the fits of cTFIM.)}

For the post-measurement state $\ket{\Psi^x_\pm}$ in Sec. \ref{sec:post-selection_EE}, we apply the same Kraus operator on each site to the infinite chain and fit $S$ to $\log \xi$ to obtain the effective central charge $c_\text{eff}$. The so-obtained effective central charge is presented in Fig.\ref{fig:ceff1} and is in perfect agreement with the results obtained using the alternative numerical method based on the Majorana-fermion representation of the spin chain discussed in App. \ref{app:Majorana}.

For the Born-rule measurements and forced measurements, the Kraus operators are applied to the chain according to the corresponding probabilities in a periodic fashion whose unit cell is of a fixed length $L$. For each bond within the unit cell, we calculate the half-chain entanglement entropy. The effective central charge is obtained by fitting the average EE over the bonds to $\log \xi$. We then perform the one-by-one measurement procedure ten times to get the standard deviation of the effective central charge.

\section{Analytic calculation of the EE of post-measurement states $\ket{\Psi^x_\pm}$ and mapping to the single-bond defect TFIM}
\label{app:lattice}

We calculate the half-system EE of the post-measurement states $\ket{\Psi^x_\pm}$ with spatially uniform measurement outcomes $\{\m_j=+\}$ and $\{\m_j=-\}$. This calculation is a generalization of Ref. \onlinecite{eisler2010}. We take half-system EE of $\ket{\Psi^x_+}$ as an example. The analysis for $\ket{\Psi^x_-}$ is completely parallel.

\begin{figure}[hp]
    \centering
    \captionsetup{justification=RaggedRight}
    \subfloat[\label{fig:cut}]{\includegraphics[width = .45\linewidth]{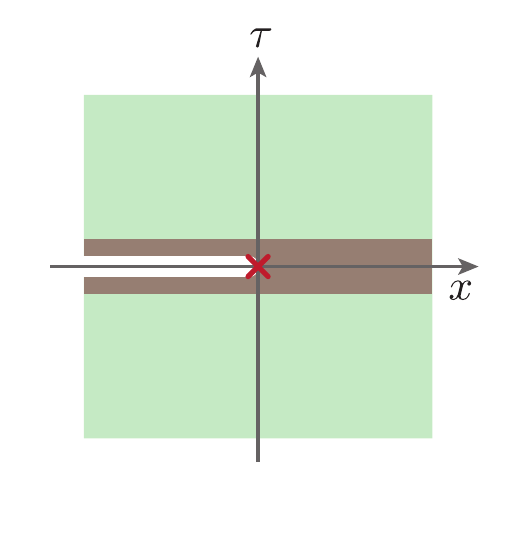}}
    \subfloat[\label{fig:expmap}]{\includegraphics[width = .45\linewidth]{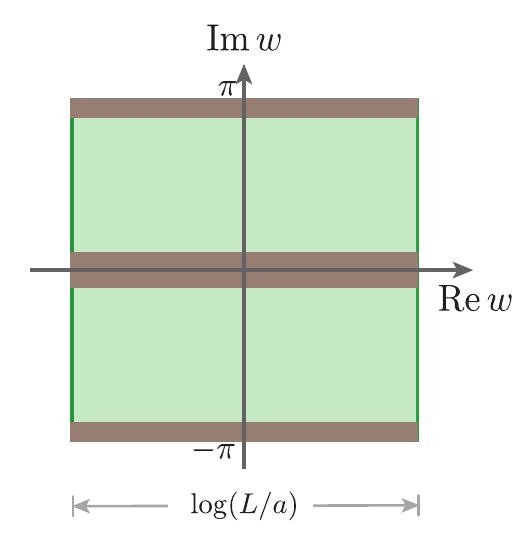}}
    \caption{half-system reduced density matrix $\rho_A$ of $\ket{\Psi_+^x}$ represented as a path integral in (a) infinite plane geometry, (b) rectangular geometry after a conformal map. The green patches are described by the Ising CFT. The brown patches are where the measurement-induced perturbation locates. }\label{fig:cuts}
\end{figure}

The post-measurement state $\ket{\Psi^x_+}$ is given by
\begin{align}
    \ket{\Psi^x_+} = \frac{\prod_j K^x_{j,+}\ket{\Omega} }{|| \prod_j K^x_{j,+}\ket{\Omega} ||} \propto e^{\beta \sum_j \sigma^x_j} \IsGS,
\end{align}
In the continuum limit and $L\rightarrow\infty$, the state is described by a path integral on a half Euclidean plane with Ising Lagrangian and a perturbation at $\tau =0$. Operator expectations and correlations are calculated by path integral on a full plane with operator insertions on $\tau=0$ slice. The unnormalized half-system (for interval $A = (-\infty, 0)$) density matrix $\rho_A$ is obtained by taking two copies of such path integrals (upper half plane $\bra{\Psi^x_+}$ and lower half plane $\ket{\Psi^x_+}$) and gluing along the half line $B = (0, \infty)$ (Fig. \ref{fig:cut}). Our goal is to calculate $\rho_A$. First, we apply a conformal transformation,
\begin{equation}
w = \log(z)
\end{equation}
where $z = x+ i \tau$. 

\begin{figure}[h!]
    \centering
    \captionsetup{justification = RaggedRight}
    \subfloat[\label{fig:exp2}]{\includegraphics[width = .5\linewidth]{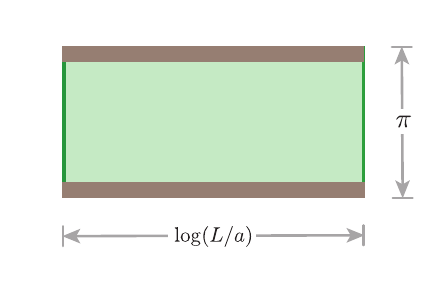}}
    \subfloat[\label{fig:lattice}]{\includegraphics[width = .5\linewidth]{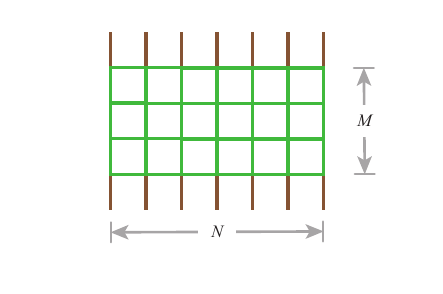}}
    \caption{$\rho_A^{1/2}$ as (a) Ising CFT path integral with perturbation on the edges, (b) isotropic critical classical Ising model with line defects. The green bonds are $K_c$. The brown bonds are $K_0'$.}\label{fig:rho1d2}
\end{figure}

The geometry in $w$ is shown in Fig. \ref{fig:expmap}. This geometry consists of two identical pieces stacked together. It suffices to calculate half of it $\rho_A^{1/2}$ (Fig. \ref{fig:exp2}). We now map the quantum cTFIM to a classical Ising model in 2d. In doing so, we effectively discretize both space and time (Fig. \ref{fig:lattice}). 

In operator language, the path integral on the full plane (Fig. \ref{fig:cut}) is 
\begin{equation}
Z_\text{q} = \lim_{T\rightarrow \infty} \Tr\left( e^{-T\hat{H}}e^{2\beta\sum_j\sigma_j^x}e^{-T\hat{H}} \right) \label{eq:zq}
\end{equation}
We now map this to a classical Ising model in 2d with a line defect (for a review, see Ref. \onlinecite{igloi1993}). Generically, the classical Hamiltonian is,
\begin{equation}
\begin{split}
  H_\text{cl} =  &- \sum_{\vec{r}}( K_1 \sigma_{\vec{r}}\sigma_{\vec{r}+\hat{x}} 
   + K_2 \sigma_{\vec{r}} \sigma_{\vec{r}+\hat{y}} )  \\& ~~~~~ - \sum_x (K_0-K_2)\sigma_{(x, y=0)}\sigma_{(x, y=1)}, 
\end{split} \label{eq:qz}
\end{equation}
where $\sigma_{\vec r} = \pm 1$ is the classical Ising spin at site $\vec r = (x,y)$. We put the Hamiltonian on an $L\times L$ lattice so that $-L/2<x,y\leq L/2$. The bulk horizontal and vertical couplings are $K_1$ and $K_2$, while on the line defect the vertical bond is modified to $K_0$. The classical partition function is,
\begin{align}
Z_\text{cl} = e^{-H_\text{cl}} = \Tr\left(u^{L/2} e^{(K_0^* - K_2^*)\sum_j \sigma^x_j}u^{L/2}\right) \label{eq:clz}
\end{align}
with a transfer matrix $u$ defined by
\begin{align}
u = e^{K_2^*\sum_j \sigma_j^x} e^{K_1\sum_j \sigma_j^z\sigma_{j+1}^z} \label{eq:tm}
\end{align}
Here, we define the dual couplings $K_0^*$ and $K_2^*$ via
\begin{align}
K^* \equiv -\frac{1}{2}\log(\tanh(K)) \label{eq:dual}
\end{align} 
To match the classical and quantum partition functions, we need to take the Hamiltonian limit, $K_1\rightarrow 0$, $K_2^*\rightarrow 0$, and make the identifications,
\begin{align}
& K_1 = K_2^* \nonumber \\
& K_1 \frac{L}{2}  = T \label{eq:map_q_cl}\\
& 2\beta = K_0^* - K_2^* 
\nonumber 
\end{align}
Since we want to harness conformal symmetries shortly, we can further substitute this classical Ising model in the Hamiltonian limit 
with another critical isotropic Ising model (deformed by a defect), where the horizontal and vertical bonds are of the same and finite strength. These two classical models share the same IR behavior, as explained below. The classical Hamiltonian of the latter model is given by
\begin{equation}
\begin{split}
 H_\text{iso} =  &- \sum_{\vec{r}}( K_c \sigma_{\vec{r}}\sigma_{\vec{r}+\hat{x}} 
   + K_c \sigma_{\vec{r}} \sigma_{\vec{r}+\hat{y}} ) \\&~~~~~- \sum_x (K_0'-K_c)\sigma_{(x, y=0)}\sigma_{(x, y=1)}
   \end{split}\label{eq:hiso}
\end{equation}
where $K_c =K_c^*= \text{arctanh}(\sqrt{2}-1)$, which ensures the criticality in the bulk, and the defect strength $K_0'\neq K_0$ is to be determined shortly. Without defects, the IR properties of all classical Ising models $K_1 = K_2^*$ are the same at criticality.
 With the defect line, we need to match the critical exponent \cite{bariev1979, mccoy1980}
\begin{equation}
    \Delta_{z,{\rm ladder}} = \frac{2}{\pi^2}\arctan^2\left(e^{2(K_0^* - K_2^*)}\right).
    \label{eq:Delta_z_defect}
\end{equation}
Here, $\Delta_{z,{\rm ladder}}$ is defined by, 
\begin{equation}
\braket{\sigma_{\vec{r}} \sigma_{\vec{r}+\vec{r}_\parallel}} \sim |\vec{r}_\parallel|^{-2\Delta_{z,{\rm ladder}}}
\end{equation}
for $|\vec{r}_\parallel|\gg 1$ and $\vec{r}_\parallel$ parallel to the defect line (which is the $x$-direction in the cases above). The defect bonds form the shape of a ladder, thus the name. Equivalently, $\Delta_{z,{\rm ladder}}$ can be defined on a quantum state as in section \ref{sec:post-selection_EE}. 
Matching $\Delta_{z,{\rm ladder}}$ for $H_\text{cl}$ and $H_\text{iso}$, we obtain,
\begin{equation}
K_0^* - K_2^* = 2\beta = K_0'^* - K_c^*. \label{eq:beta}
\end{equation}
Combining this identification with Eq. \eqref{eq:Delta_z_defect}, we obtain the expression of $\Delta_{z,+}$ in Eq. \eqref{eq:DeltaZ_post2}. $\Delta_{z,-}$ can be obtained by simply substituting $\beta\rightarrow -\beta$. 

We can now calculate the half-system EE in the isotropic model Eq. \eqref{eq:hiso}. Put this model on an $M\times N$ lattice (Fig. \ref{fig:lattice}), where each of the top and bottom ladders (brown) includes half of the defect bond $K_0'$. Viewing this as a matrix that rotates the states on the bottom row to the states on the top row, we can immediately readout,
\begin{align}
&\rho_A^{1/2} = u_0 u^M u_0 \label{eq:rhohalf} 
\end{align}
with 
\begin{align}
&  u = e^{\frac{1}{2}K_c\sum_j \sigma^x_j}e^{K_c\sum_j\sigma^z_j \sigma^z_{j+1}}e^{\frac{1}{2}K_c\sum_j \sigma^x_j},\\
& u_0 = e^{\frac{1}{2}(K_0'^* - K_c)\sum_j \sigma^x_j} = e^{\beta\sum_j \sigma^x_j},
\end{align}
where the transfer matrix $u$ is analogous to Eq. \eqref{eq:tm}, albeit symmetrized, and in the last equality, we have used Eq. \eqref{eq:beta}. Now, we perform Jordan Wigner transformation,
\begin{align}
\sigma_j^x &= 2c^\dagger_jc_j-1 \\
\sigma_j^z \sigma_{j+1}^z &= (c^\dagger_j  - c_j)(c^\dagger_{j+1}+c_{j+1})
\end{align}
where $c_j$ are fermionic operators. It remains to bring the density matrix to the factorized form,
\begin{align}
\rho_A &= e^{-H_\text{eff}}
\end{align}
with 
\begin{align}
 H_\text{eff} = 
\sum_l\epsilon_l \tilde{c}^\dagger_l \tilde{c}_l.
\end{align}
Where the effective Hamiltonian is a sum of single fermion modes $\tilde{c}_l$ which are related to the real space fermion modes $c_j$ by some unitary transformations. Once in this form, we can calculate EE,
\begin{align}
S_{|\Psi_+^v\rangle}(L/2) &=  - \Tr_A \left(\frac{\rho_A}{\Tr_A(\rho_A)}\log\left(\frac{\rho_A}{\Tr_A(\rho_A)}\right)\right) \nonumber \\
&=\sum_l \log(1+e^{-\epsilon_l}) + \sum_l \frac{\epsilon_l}{e^{\epsilon_l}+1}
\end{align}
These last two technical steps are carried out in Ref. \onlinecite{eisler2010} which deals with the half-system EE of a cTFIM with a bond defect in the middle. We expect that the two systems share the same universal behavior of EE because they are related by a 90 degrees spacetime rotation (see Fig. \ref{fig:spacetime_rotation}). More precisely, the cTFIM Hamiltonian with the bond defect is,
\begin{equation}
\hat{H}_\text{def} = -\sum_j (\sigma^z_j \sigma^z_{j+1} + \sigma^x_j) - (t-1)\sigma^z_0\sigma^z_1
\end{equation}
where $-L/2<j\leq L/2$. The corresponding classical theory is, 
\begin{equation}
\begin{split}
  H_\text{def,cl} =  &- \sum_{\vec{r}}( K_1 \sigma_{\vec{r}}\sigma_{\vec{r}+\hat{x}} 
   + K_2 \sigma_{\vec{r}} \sigma_{\vec{r}+\hat{y}} )  \\& ~~~~~ - \sum_y (t-1)K_1\sigma_{(x=0, y)}\sigma_{(x=1, y)}.
\end{split} \label{eq:defcl}
\end{equation}
Comparing Eqs. \eqref{eq:qz} and \eqref{eq:defcl} , we see that the two geometries are related by a 90 degrees rotation. To relate the defect couplings $K_0$ and $tK_1$, we rescale $H_\text{def,cl}$ to the (rotated) isotropic theory Eq. \eqref{eq:hiso} by matching the critical exponent $\Delta_{z,{\rm ladder}}$, 
\begin{equation}
\frac{\tanh(K_1)}{\tanh(tK_1)} = \tan\left(\sqrt{\frac{\pi^2\Delta_{z,\text{ladder}}}{2}}\right)= e^{2(K_0'^* - K_c)}
\end{equation}
where in the first equality we have used Eq. \eqref{eq:Delta_z_defect} with $K_1$ and $K_2$ exchanged and the identity,
\begin{equation}
e^{-2K^*} \equiv \tanh(K).
\end{equation}
In the Hamiltonian limit $K_1\rightarrow 0$, thus we identify, 
\begin{equation}
t = e^{-2(K_0'^* - K_c)} = e^{-4\beta} \label{eq:tbeta}
\end{equation}
where in the last equality we have used Eq. \eqref{eq:beta}.

Comparing Fig. \ref{fig:lattice} with Fig. 4 in Ref. \onlinecite{eisler2010}, we see that the two expressions of $\rho_A^{1/2}$ have the same form, up to a cyclic rotation of the matrices. The cyclic rotation does not change $\Tr(\rho_A^n)$ for any $n$, thus it also preserves EE. We then get the final formulae of EE (Eqs. \eqref{eq:EE_log_scaling_x_measured} and \eqref{eq:c_eff}) using Eqs. (22) and (26) of Ref \onlinecite{eisler2010} and identifying $s=\frac{2}{t+\frac{1}{t}}=\frac{1}{\cosh(4\beta)}$ using Eq. \eqref{eq:tbeta}. 

Finally, let us point out that the two cases $\ket{\Psi_+^x}$ and $\ket{\Psi_-^x}$ are related by Kramers-Wannier (KW) duality. To see this, let us first look at the quantum duality transformation $U_D$,
\begin{align}
U^\dagger_D \sigma^x_j U_D  &= \sigma^z_{j-1}\sigma^z_{j} \nonumber \\
U^\dagger_D \sigma^z_j \sigma^z_{j+1} U_D &= \sigma^x_j.
\end{align}
The critical ground state $\ket{\Omega}$ is invariant under $U_D$ since it is a symmetry of the critical Hamiltonian Eq. \eqref{eq:TFIM}. Apply this to the state $\ket{\Psi^x_-}$,
\begin{align}
U_D^\dagger \ket{\Psi^x_-} &\propto U_D^\dagger e^{-\beta \sum_j \sigma^x_j}\ket{\Omega} \nonumber\\
&= e^{-\beta \sum_j \sigma_{j-1}^z\sigma_j^z }\ket{\Omega} \nonumber\\
&= e^{-\beta \sum_j \sigma_{j-1}^z\sigma_j^z} e^{-\beta \hat{H}}e^{\beta \hat{H}}\ket{\Omega} \nonumber\\
&\approx e^{\beta E_0}e^{\beta\sum_j \sigma^x_j}\ket{\Omega} \propto \ket{\Psi^x_+}
\end{align}
where $\hat{H}$ is the critical Hamiltonian Eq. \eqref{eq:TFIM} and the last equality holds only if $\beta\ll1$. For finite $\beta$, we can relate the two cases by classical KW duality as follows. Starting with the classical representation of $\ket{\Psi^x_+}$ Eq. \eqref{eq:qz}. This is a critical 2d classical Ising model with a ladder defect along the line $y=0$ and defect coupling $K_0$. The classical KW transforms the lattice onto its dual lattice, which for square lattice is again a square. The coupling $K$ on every link of the original lattices is dualized to $K^*$ according to Eq. \eqref{eq:dual} and put on the corresponding link of the dual lattice. For a square lattice, a horizontal link in the original lattice is mapped to a vertical link in the dual lattice and vice versa. Thus the bulk horizontal and vertical couplings of the dual theory $\tilde{K}_1$ and $\tilde{K}_2$ reads,
\begin{align}
    \tilde{K}_1 &= K_2^*=K_1\\
    \tilde{K}_2 &= K_1^*=K_2
\end{align}
We see the bulk coupling is invariant at criticality. The ladder defect becomes a chain defect (horizontal defect bonds along $y=0$) in the dual theory with couplings,
\begin{equation}
\tilde{K_0}  = K_0^*
\end{equation}
Both the chain and ladder defects look the same in the IR. So by matching the critical exponents, we can substitute the chain defect with a ladder defect with a different coupling $K_0^\vee$. The critical exponent for the chain defect reads \cite{bariev1979, mccoy1980},
\begin{align}
\Delta_{z,\text{chain}} &= \frac{2}{\pi^2}\arctan^2\left(e^{-2(\tilde{K}_0-\tilde{K}_1)} \right) \nonumber \\
&=\frac{2}{\pi^2}\arctan^2\left(e^{-2(K_0^*-K_2^*)} \right)
\end{align}
while applying Eq. \eqref{eq:Delta_z_defect} to the new ladder defect $K_0^\vee$
\begin{equation}
\Delta_{z, \text{ladder}} = \frac{2}{\pi^2}\arctan^2\left(e^{2(K_0^{\vee*} - K_2^*)}\right)
\end{equation}
Equating the two critical exponents and using Eq. \eqref{eq:map_q_cl}, we finally get,
\begin{equation}
K_0^{\vee*} - K_2^* = - (K_0^*-K_2^*)  = -2\beta
\end{equation}
Mapping back to the quantum theory (c.f. Eqs. \eqref{eq:zq},\eqref{eq:clz}), we see that the new ladder theory simply replaces $\beta$ with $-\beta$ and indeed corresponds to $\ket{\Psi^x_-}$. Since the KW duality preserves the partition function, we expect that the entanglement entropy is also preserved, being a replicated and twisted generalization of the partition function. This explains why $\Ceff$, which depends only on $\cosh(4\beta)$ is insensitive to the sign of $\beta$.

\section{Numerical simulation in a non-interacting Majorana-fermion representation}

\label{app:Majorana}
The cTFIM Hamiltonian can be mapped, under the Jordan-Wigner transformation, to a non-interacting fermion Hamiltonian on a one-dimensional Majorana-fermion chain:
\begin{align}
    H_{\rm maj} = -\sum_j \i\hgamma_j \hgamma_{j+1} 
    \label{eq:Maj_Ham}
\end{align}
where $\hgamma_j$'s are the Majorana-fermion operators. The spin operators in the cTFIM can be identified as follows
\begin{align}
    \sigma_j^x = \i \hgamma_{2j-1} \hgamma_{2j},~~~~\sigma_j^z \sigma_{j+1}^z = \i \hgamma_{2j} \hgamma_{2j+1}.
\end{align}
The ground state $\IsGS$ of the cTFIM can be viewed as the ground state of a non-interacting-fermion Hamiltonian Eq. \eqref{eq:Maj_Ham}. 

Any non-interacting fermion state, including 
$\IsGS$, can be fully described by its covariance matrix 
\begin{align}
    \Gamma_{jj'} = \frac{\i}{2} \big\langle  [\hgamma_j,\hgamma_{j'}] \big\rangle
\end{align}
which encodes all the Majorana-fermion two-point correlation functions in the state. All the multi-point Majorana-fermion correlation functions can be calculated from the two-point functions using Wick's theorem. 

In the fermion language, a weak measurement of $\sigma_j^x$ is equivalent to a weak measurement of the observable $\i \hgamma_{2j-1} \hgamma_{2j}$. For any non-interacting fermion state $|\psi\rangle$ whose covariance matrix is given by $\Gamma$, the post-measurement state 
\begin{align}
    |\psi'\rangle = \frac{K^x_{j,\m_j} |\psi \rangle }{||  K^x_{j,\m_j} | \psi \rangle ||}
\end{align}
is still a non-interacting fermion state, which can be captured by its covariance matrix $\Gamma'$:
\begin{widetext}
\begin{align*}
  \Gamma'& = \left( \begin{array}{cccc}
        \Gamma_{[1,2j-2],[1,2j-2]} & 0 &  \Gamma_{[1,2j-2],[2j+1,2L]}\\
       0 &  -\i n_1 \sigma^y & 0\\
        \Gamma_{[2j+1,2L],[1,2j-2]}  & 0 &  \Gamma_{[2j+1,2L], [2j+1,2L]}
    \end{array}  
    \right)
   \nonumber \\
   & -   
    \left( \begin{array}{cc}
        \Gamma_{[1,2j-2],[2j-1,2j]}   & 0 \\
      0  &    -n_2 \openone \\
          \Gamma_{[2j+1,2L],[2j-1,2j]}   &  0 
    \end{array}  
    \right).
       \left( \begin{array}{cc}
        \Gamma_{[2j-1,2j],[2j-1,2j]} &  \openone \\
        - \openone &   \i n_1 \sigma^y \\
    \end{array}  
    \right)^{-1}.
     \left( \begin{array}{ccc}
        \Gamma_{[2j-1,2j],[1,2j-2]}   & 0 &   \Gamma_{[2j-1,2j],[2j+1,2L]} \\
      0  &    n_2 \openone   & 0
    \end{array}  
    \right),
\end{align*}
\end{widetext}
where $\Gamma_{[j_1,j_2], [j_1',j_2']}$ denotes the block of the covariance matrix $\Gamma$ with the rows ranging from $j_1$ to $j_2$ and the columns ranging from $j_1'$ to $j_2'$. Here, we've assumed that there are $2L$ Majorana modes $\hgamma_{j=1,...,2L}$ in the one-dimensional chain. The parameters $n_1$ and $n_2$ are determined by the measurement strength $\lambda$ and the outcome $\m_j$:
\begin{align}
    n_1 = \frac{-2\m_j\lambda}{1+\lambda^2},~~~~n_2 = \frac{1-\lambda^2}{1+\lambda^2}.
\end{align}
The Born-rule probability Eq. \eqref{eq:Born-ruleX} can also be expressed using the non-interacting fermion representation
\begin{align}
     p^x(\m_j) &= \langle \psi | (K^x_{j,\m_j})^\dag K^x_{j,\m_j}  |\psi \rangle 
= \frac{1+\lambda^2 + 2\m_j \lambda \Gamma_{2j-1,2j}}{2(1+\lambda^2)}.
\end{align}

Using the covariance matrix formulation, we can perform efficient numerical calculations of the post-measurement non-interacting fermion states resulting from the weak measurements along the spin-$x$ axis on the cTFIM ground state $\IsGS$. 

Given the covariance matrix $\Gamma$ of a non-interacting fermion state, one can directly calculate the subsystem von Neumann EE. For example, for a subsystem that is an interval starting from the $j$th Majorana-fermion site and ending on the $j'$th site, the subsystem EE, is given by
\begin{align}
&S_{[j,j']} \nonumber \\
&
= -\frac{1}{2 }\sum_{s=\pm 1}\Tr\left( \frac{\openone + s \i \ \Gamma_{[j,j'], [j,j']} }{2} \log \frac{\openone + s \i \ \Gamma_{[j,j'], [j,j']} }{2}\right).   
\end{align}
The connected spin-$x$ correlation function $\langle \sigma_j^x \sigma_{j'}^x \rangle - \langle \sigma_j^x \rangle \langle\sigma_{j'}^x \rangle$ can be written as
\begin{align}
   &\langle \sigma_j^x \sigma_{j'}^x \rangle - \langle \sigma_j^x \rangle \langle\sigma_{j'}^x \rangle 
   \nonumber \\ 
   & ~~~ =  -\Gamma_{[2j-1,2j'-1]}\Gamma_{[2j,2j']}+\Gamma_{[2j-1,2j']}\Gamma_{[2j,2j'-1]}.
\end{align}
The absolute value of the spin-$z$ correlation function $| \langle \sigma_j^z \sigma_{j'}^z \rangle|$ can also be calculated in the covariance matrix formalism. Assuming $j'>j$, we can write
\begin{align}
    \sigma_j^z \sigma_{j'}^z = (\i)^{j'-j}
    \prod_{2j\leq k \leq 2j'-1} \hgamma_k.
\end{align}
For a non-interacting fermion state $|\psi\rangle$ with covariance matrix $\Gamma$, the state $|\psi'\rangle = \sigma_j^z \sigma_{j'}^z|\psi\rangle$ is also a non-interacting fermion state. Its covariance matrix is given by 
\begin{align}
    \Gamma' = \Lambda_{[2j,2j'-1]} \cdot\Gamma\cdot \Lambda_{[2j,2j'-1]},
\end{align}
where $\Lambda_{[2j,2j'-1]}$ is a diagonal matrix with $-1$ between (and including) the $(2j)$th and the $(2j'-1)$th entries, and $1$ for all other entries. The absolute value of the spin-$z$ correlation function can be written as
\begin{align}
   |\langle \sigma_j^z \sigma_{j'}^z \rangle| = |\langle \psi' | \psi \rangle| =  \Big| \frac{1}{2^{2L}} {\rm Pf} \left( \begin{array}{cc}  \i \Gamma &  \openone \\ - \openone &  \i \Gamma'\end{array} \right) \Big|^{\frac{1}{2}},
\end{align}
where $2L$ is the number of sites in the Majorana-fermion chain (and $L$ is the number of sites in the corresponding spin chain). Here, in the second equality, we have applied the result from Ref. \onlinecite{Bravyi2004FermionicOptics}.

\section{Optimal biased forced measurement as a mean-field approximation of the Born-rule measurement}
\label{app:OptimalBias}
In the biased forced measurements, the measurement outcomes on each site are weighted by probability $p_{\pm}$. If we view the Born-rule measurement and the biased forced measurement as different probability distributions over the measurement outcomes $\{\m_j = \pm\}$, we would like to ask what is the ``closest" biased forced measurement compared to the Born-rule one. To address this question, we calculate the Kullback–Leibler(KL) divergence (relative entropy)\cite{KL1951} of the force measurement w.r.t the Born-rule one,
\begin{equation}
    D_{KL}(p||p') \equiv \sum_{\{\m_j\}} p(\{\m_j\}) \log\left[\frac{p(\{\m_j\})}{p'(\{\m_j\})}\right],
\end{equation}
where $p(\{\m_j\})$ is the Born-rule probablity given in Eq. \eqref{eq:Born-Rule_Chain} and the $p'(\{\m_j\})$ is the probablity distribution Eq. \eqref{eq:biased_forced_prob} for the biased forced measurement. Plugging in the Eq. \eqref{eq:biased_forced_prob}, we can write
\begin{equation}
\begin{split}
    &D_{KL}(p||p') \\=& \sum_{\{\m_j\}} p(\{\m_j\}) \log~p(\{\m_j\})\\
    &- \sum_{m=0}^L P_m \left[ m ~\log~p_+ + (L-m) \log(1-p_+)\right],
\end{split}
\end{equation}
where $P_m$ is the sum of $p(\{\m_j\})$ with $m = N_+(\{\m_j\})$.

As a function of $p_+$, $D_{KL}(p||p')$ is minimized by $p_+^* = P^{(1)}/N$ and
\begin{equation}
    P^{(1)} \equiv \sum_{m=0}^N m~P_m.
\end{equation}
We evaluate $P^{(1)}$ by considering a generating function $\Sigma(t)$ defined by
\begin{equation}
    \begin{split}
    \Sigma(t) &\equiv \Big\langle \prod_{i=1}^N \left[\left(1+\frac{t}{2}\right)K_{i,+}^2+ \left(1-\frac{t}{2}\right) K_{i,-}^2\right] \Big\rangle\\
    & = \sum_{m = 0}^N \left(1+\frac{t}{2}\right)^m \left(1-\frac{t}{2}\right)^{N-m} P_m.
    \end{split}
\end{equation}
Using this generating function, we can write
\begin{equation}
    \frac{P^{(1)}}{N} = \frac12 + \frac{1}{N}\Sigma'(t=0)
\end{equation}

On the other hand, by plugging in the definition of $K_{i,\pm}$'s, we have
\begin{equation}
   \Sigma(t) = (\cosh{\beta'})^{-N} \langle e^{\beta' \sum_i \sigma^x_i} \rangle, 
\end{equation}
where $\tanh{\beta'} = \frac{\lambda t}{1+\lambda^2}$.

It is readily seen that $p_+^* = \frac{1}{2} + \frac{\lambda}{1+\lambda^2} M_x$, where $M_x = \frac{1}{N} \sum_i \langle \sigma^x_i \rangle = \frac{2}{\pi}$ is the total magnetization density along $x$-direction in the cTFIM ground state $\IsGS$.

\section{Analytic study of the effective central charge for Born-rule and forced measurement}
\label{app:ceff}
In this section, we analytically study the behavior of the effective central charge $c_{\rm eff}$ in the three cases with Born-rule measurements, unbiased forced measurements, and biased forced measurements. The key to our analysis is finding the approximations of the measurement-induced perturbations that capture their most dominant effects under RG. In all three cases, the approximated version of the measurement-induced perturbations turns out to be of the same type as the perturbation occurring in the post-measurement states $\ket{\Psi^x_{\{\m_j = \pm \}} }$ which we have carefully studied in Sec. \ref{sec:post-selection_EE}. Using this approximation, we obtain the expression of the effective central charge $c_{\rm eff}$ in the three cases with the Born-rule measurements, the unbiased forced measurements, and biased forced measurements.

\subsection{Born-rule and unbiased forced measurements}
\label{app:ceff_b_ubf}
In the cases with the Born-rule and the unbiased forced measurements, the effect of the measurements is captured by the perturbation Eq. \eqref{eq:deltaS_outcome_summed} in the continuum. The two types of measurements correspond to different replica limits: $R\rightarrow 1$ for Born-rule measurements and $R\rightarrow 0$ for unbiased forced measurements.

At the microscopic level, this measurement-induced perturbation Eq. \eqref{eq:deltaS_outcome_summed} is given by the operator 
\begin{align}
   {\cal M}_R & = \sum_{\{\m_j = \pm \}} \exp\left( 
2\beta\sum_j \m_j \sum_{\alpha=1}^R \sigma^x_{j,\alpha}\right) 
\nonumber \\
&= \prod_j 2 \cosh\left(2\beta\sum_{\alpha=1}^R\sigma^x_{j,\alpha}\right)
 \label{eq:ave_Measurement_Op}
\end{align}
acting on $R$ replicas of the Ising spin chain. $\lambda = \tanh{\beta}$ is measurement strength. $\sigma^x_{j,\alpha}$ is the Pauli operator that acts on the $j$th site in the $\alpha$th replica of the Ising spin chain. For each $j$, we can write  
\begin{align}
    &\cosh\left(2\beta\sum_{\alpha=1}^R \sigma^x_{\alpha}\right) 
    \nonumber \\
    & ~~~~~~ = \sum_{r=0}^{\lfloor \frac{R}{2}\rfloor} \mathtt{c}^{R-2r}\mathtt{s}^{2r}\sum_{\alpha_1<\alpha_2<\ldots<\alpha_{2r}} \sigma^x_{\alpha_1}\sigma^x_{\alpha_2}\cdots\sigma^x_{\alpha_{2r}}
    \label{eq:Cosh_Expansion}
\end{align}
where we've used the shorthand notations $\mathtt{c}\equiv \cosh(2\beta)$, $\mathtt{s}\equiv \sinh(2\beta)$ and $\mathtt{t}\equiv \tanh(2\beta)$. The site index $j$ has been suppressed. When we consider the action of each $\sigma^x$ operator on the cTFIM ground state $\ket{\Omega}$, we can decompose each $\sigma^x$ as
\begin{align}
    \sigma^x = \langle \sigma^x \rangle + \phi^x
\end{align}
where $\langle \sigma^x \rangle \equiv \langle \Omega | \sigma^x_j \IsGS = 2/\pi $ and $\phi^x$ corresponds to the energy field of the Ising CFT (with scaling dimension 1). We expand the product $\prod_i \sigma^x_{\alpha_i}$ appearing in Eq. \eqref{eq:Cosh_Expansion} as
\begin{align}
    \prod_{i=1}^{2r} \sigma^x_{\alpha_i} = \langle \sigma^x \rangle^{2r} + \langle \sigma^x \rangle^{2r-1} \sum_{i=1}^{2r} \phi^x_{\alpha_i} + O\left((\phi^x)^2\right)
    \label{eq:Sigma_Prod_expansion}
\end{align}
to the first order of $\phi^x_{\alpha_i}$'s. Recall that, from the spacetime perspective, the measurement-induced perturbation occurs on a one-dimensional defect at imaginary time $\tau = 0$. We keep the first-order terms in $\phi^x_{\alpha_i}$'s because they give rise to marginal perturbations to the ($R$-replica) Ising CFT, while the higher-order terms are irrelevant under RG. Using the expansion above, we can write 
\begin{align}
    &\cosh\left(2\beta\sum_\alpha \sigma^x_{\alpha}\right) = \mathtt{c}^R \frac{(1+x)^R+(1-x)^R}{2}  \nonumber \\
    &  ~~~~~~ + \mathtt{c}^R \mathtt{t} \frac{(1+x)^{R-1}-(1-x)^{R-1}}{2} \sum_{\alpha=1}^R \phi^x_\alpha +   O\left((\phi^x)^2\right),
\end{align}
where $x \equiv \mathtt{t} \langle \sigma^x \rangle$. To the linear order of $\phi^x$, we can approximate $\cosh\left(2\beta\sum_\alpha \sigma^x_{\alpha}\right)$ by
\begin{align}
    \cosh\left(2\beta\sum_\alpha \sigma^x_{\alpha}\right) = \tilde{A} \prod_{\alpha = 1}^R \left(1+\tilde{g} \sigma^x_\alpha \right) +  O\left((\phi^x)^2\right)
    \label{eq:cosh_approx}
\end{align}
Here, the coefficients $\tilde{A}$ and $\tilde{g}$, which depend on both the measurement strength $\lambda = \tanh \beta$ and the replica number $R$, are given by 
\begin{align}
\tilde{g} &= \frac{g}{1+(R-1)\braket{\sigma^x}g} \label{eq:ren1}\\
\tilde{A} &= A\frac{[1+(R-1)\braket{\sigma^x}g]^R}{[1+R\braket{\sigma^x}g]^{R-1}} \label{eq:ren2}
\end{align}
where
\begin{align}
    &g = \frac{\mathtt{t}\left[1-\left(\frac{1-x}{1+x}\right)^{R-1}\right]}{[1-(R-1)x]+[1+(R-1)x]\left(\frac{1-x}{1+x}\right)^{R-1}}, \\
    &A = 
    \nonumber \\
    &\mathtt{c}^R\frac{(1-x)^{R-1}[(R-1)x+1]-(1+x)^{R-1}[(R-1)x-1] }{2}.
\end{align}

Now, we can write down an approximated version of the measurement-induced perturbation ${\cal M}_R$ in Eq. \eqref{eq:ave_Measurement_Op} with the Born-rule and the unbiased forced measurements:
\begin{align}
    {\cal M}_R = \prod_{\alpha = 1}^R  \prod_j 2 \tilde{A} \left(1+\tilde{g} \sigma^x_{j,\alpha} \right) +  O\left((\phi^x)^2\right)
    \label{eq:MR_approx}
\end{align}
Notice that the first term on the right-hand side of this equation is exactly the same type of measurement-induced perturbation in the post-measurement states $\ket{\Psi^x_{\{\m_j = \pm \}} }$ discussed Sec. \ref{sec:post-selection_EE} but with a modified measurement strength captured by $\tilde{g}$. Recall that the measurement-induced perturbations in the post-selected states $\ket{\Psi^x_{\{\m_j = \pm \}} }$ are exactly marginal under RG. The approximation Eq. \eqref{eq:MR_approx} includes all the marginal perturbation and neglects the terms of the order $O\left((\phi^x)^2\right)$, which are irrelevant under RG. ${\cal M}_R$ does not include any RG-relevant terms. Hence, we expect this approximation Eq. \eqref{eq:MR_approx} to capture the behavior of the measurement-induced perturbation Eq. \eqref{eq:ave_Measurement_Op} and Eq. \eqref{eq:deltaS_outcome_summed} accurately (at least) for small measurement strength. 

For the case with Born-rule measurements, we should consider the replica limit $R\rightarrow 1$. We find that
\begin{align}
    \tilde{g} \xrightarrow{R\to 1} 0,
\end{align}
which implies that the measurement-induced perturbation is irrelevant under RG. The effective central charge defined through the subsystem EE should remain the same as the ``un-measured" cTFIM ground state $|\Omega \rangle$. In other words, we conclude that, under Born-rule measurements, $c_{\rm eff} = 1/2$, which is independent of the measurement strength $\lambda$. This result is consistent with our numerical study presented in Fig. \ref{fig:born} in Sec. \ref{sec:born}.

For the case with unbiased forced measurements, we should consider the replica limit $R\rightarrow 0$. We find that
\begin{align}
    \tilde{g} \xrightarrow{R\to 0} -\mathtt{t}^2 \langle \sigma^x \rangle. 
\end{align}
The measurement-induced perturbation of the form $\prod_{\alpha}  \prod_j 2 \tilde{A} \left(1+\tilde{g} \sigma^x_{j,\alpha} \right)$ with $\tilde{g} = -\mathtt{t}^2 \langle \sigma^x \rangle$ is exactly the perturbation associated with the post-measurement state $\ket{\Psi^x_{\{\m_j = - \}} }$ with a measurement strength
\begin{align}
    \tilde{\lambda} & = \tanh\left(\frac{1}{2} {\rm arctanh}\left( \mathtt{t}^2 \langle \sigma^x\rangle \right) \right) 
    \nonumber \\
    & = \tanh\left(\frac{1}{2} {\rm arctanh}\left( \frac{4\lambda^2}{(1+\lambda^2)^2} \langle \sigma^x\rangle \right) \right). 
\end{align}
Consequently, the effective central charge $c_{\rm eff}^{\rm f}$ in the case of unbiased forced measurements as a function of the measurement strength $\lambda$ is given by 
\begin{align}
    c_\text{eff}^{\rm f}(\lambda) = c_\text{eff}^{\rm p.m.}(\tilde{\lambda}),
    \label{eq:c_eff_ubforced}
\end{align}
where $c_\text{eff}^{\rm p.m.}(\tilde{\lambda})$ denotes the effective central charge in the post-measurement state $\ket{\Psi^x_{\{\m_j = - \}} }$ with measurement strength $\tilde{\lambda}$. The exact expression of $c_\text{eff}^{\rm p.m.}$ is given by Eq. \eqref{eq:c_eff}. The superscripts ``f" and ``p.m." are introduced to distinguish the effective central charges appearing in different scenarios.

\begin{figure}[h!]
    \centering
    \captionsetup{justification = RaggedRight}
    \includegraphics[width=.9\linewidth]{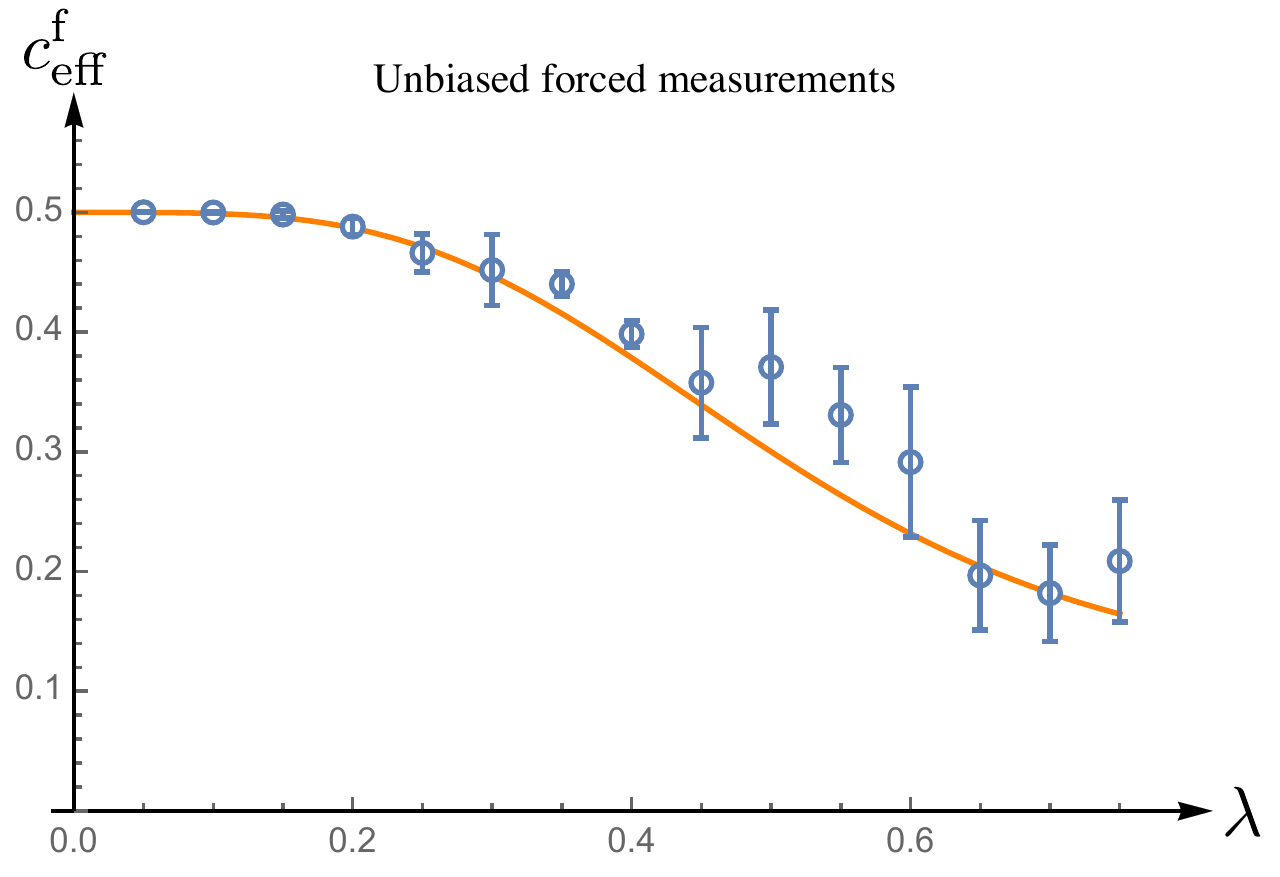}
    \caption{The effective central charge $\Ceff^{\rm f}$ of unbiased forced measurement. The orange curve is given by Eq. \eqref{eq:c_eff_ubforced}. The blue dots are the effective central charge obtained from the MPS simulations. The same numerical results are also presented in Fig. \ref{fig:forced}.}
    \label{fig:ceff_unbias}
\end{figure}

As shown in Fig. \ref{fig:ceff_unbias}, Eq. \eqref{eq:c_eff_ubforced} agrees well with the effective central charge numerically obtained from the MPS simulation of the unbiased forced measurement for measurement strength $\lambda \lesssim 0.7$. As the measurement strength $\lambda$ approaches 1, we caution that Eq. \eqref{eq:c_eff_ubforced} must receive corrections from the neglected terms in our approximation Eq. \eqref{eq:MR_approx}. We anticipate such corrections because Eq. \eqref{eq:c_eff_ubforced} does not capture the expected limit $c_\text{eff}^{\rm f}(\lambda\rightarrow 1) \rightarrow 0$. Nevertheless, our method successfully captures the behavior of the effective central charge for a large arrange of measurement strength $\lambda$.

\subsection{Biased forced measurements}
As introduced in Sec. \ref{sec:force}, with biased forced measurements, we sample each measurement outcome $\m_j = +$ with probability $p_+$ and $\m_j = -$ with probability $1-p_+$.

Under this sampling scheme, the microscopic form of the measurement-induced perturbation reads
\begin{align}
  &{\cal M}'_R = \sum_{\{\m_j = \pm \}} p_+^{N_+} (1-p_+)^{N_-} \exp\left( 
2\beta\sum_j \m_j \sum_{\alpha=1}^R \sigma^x_{j,\alpha}\right) 
\nonumber \\
& \propto \sum_{\{\m_j = \pm\}} \exp\left(\sum_j \m_j \left[\frac{1}{2}\log\left(\frac{p_+}{1-p_+}\right)+2\beta\sum_\alpha\sigma^x_{j,\alpha}\right]\right),
 \label{eq:ave_Measurement_Op_bf}  
\end{align}
with the proportionality constant $[p_+(1-p_+)]^{\frac{L}{2}}$. Here, $N_\pm$ is the number of sites where the measurement outcome $\m_j$ is $\pm$. $L$ is the total length of the Ising spin chain. The replica limit of the biased forced measurements is given by $R\rightarrow 0$. 

Similar to App. \ref{app:ceff_b_ubf}, we can approximate the measurement-induced perturbation ${\cal M}'_R$ by 
\begin{align}
   {\cal M}'_R \propto \prod_{\alpha = 1}^R  \prod_j  \left(1+\tilde{g}' \sigma^x_{j,\alpha} \right) +  O\left((\phi^x)^2\right)  
\end{align}
to the first order of $\phi^x$. In the replica limit $R\rightarrow 0$, we have 
\begin{align}
    \tilde{g}' \xrightarrow{R\to 0} \frac{2\mathtt{t} \delta p }{(1-x^2) -2x \delta p},
\end{align}
where $\delta p = p_+ - p_x^b$ is defined as the difference between $p_+$ and $p_x^b = \frac{1+x}{2} = \frac{1}{2} + \frac{\lambda}{1+\lambda^2}\braket{\sigma^x}$. The latter is the probability of measuring $\m=+1$ when we only perform the $\sigma^x$-measurement on a single site in the cTFIM ground state $\ket{\Omega}$.

When $\delta p =0$, we have $\tilde{g}' \Big|_{R \rightarrow 0} = 0$. Therefore, the corresponding effective central charge $c^{\rm bf}_{\rm eff}$ with biased forced measurements should remain the same as the un-measured cTFIM ground state $\ket{\Omega}$, i.e., $c_{\rm eff} = 1/2$, independent on the measurement strength $\lambda$. This conclusion is consistent with our numerical simulation shown in Fig. \ref{fig:forced}. Note that, when $p_+ = p_x^b$, namely $\delta p= 0$, the biased forced measurement provides a mean-field approximation to the case with the Born-rule measurement. The independence of the effective central charge on the measurement strength is found analytically and numerically in both types of measurements. 

For a generic value of $\delta p$, the effective central charge $c_{\rm eff}^{\rm bf}$ is given by
\begin{align}
        c_\text{eff}^{\rm bf}(\lambda) = c_\text{eff}^{\rm p.m.}(\tilde{\lambda}')
        \label{eq:ceff_bforced}
\end{align}
with
\begin{align}
    \tilde{\lambda}'(\delta p, \lambda) &= \tanh\left(\frac{1}{2}\text{arctanh}\left(|\tilde{g}'|\Big|_{R \rightarrow 0}\right)\right) \label{eq:bias2}. 
\end{align}
Again, this result is obtained by comparing the approximated version of the measurement-induced perturbation ${\cal M}'_R$ with the exactly marginal perturbation appearing in the post-measurement states $\ket{\Psi^x_{\{\m_j = \pm \}} }$.

\begin{figure*}[th!]
    \centering
    \captionsetup{justification = RaggedRight}
    \subfloat[]{\includegraphics[width=.4\linewidth]{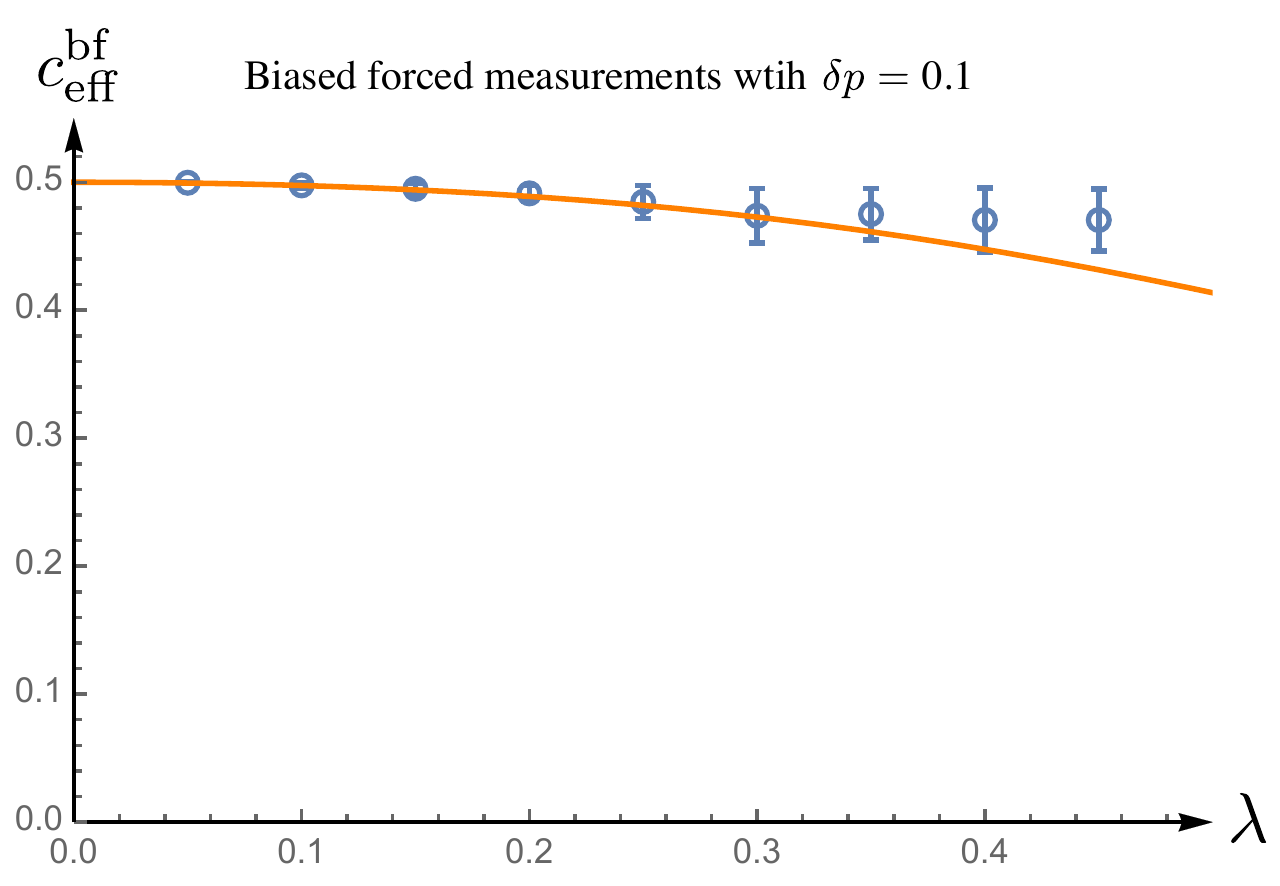}}
    \hspace{.05\linewidth}
    \subfloat[]{\includegraphics[width=.4\linewidth]{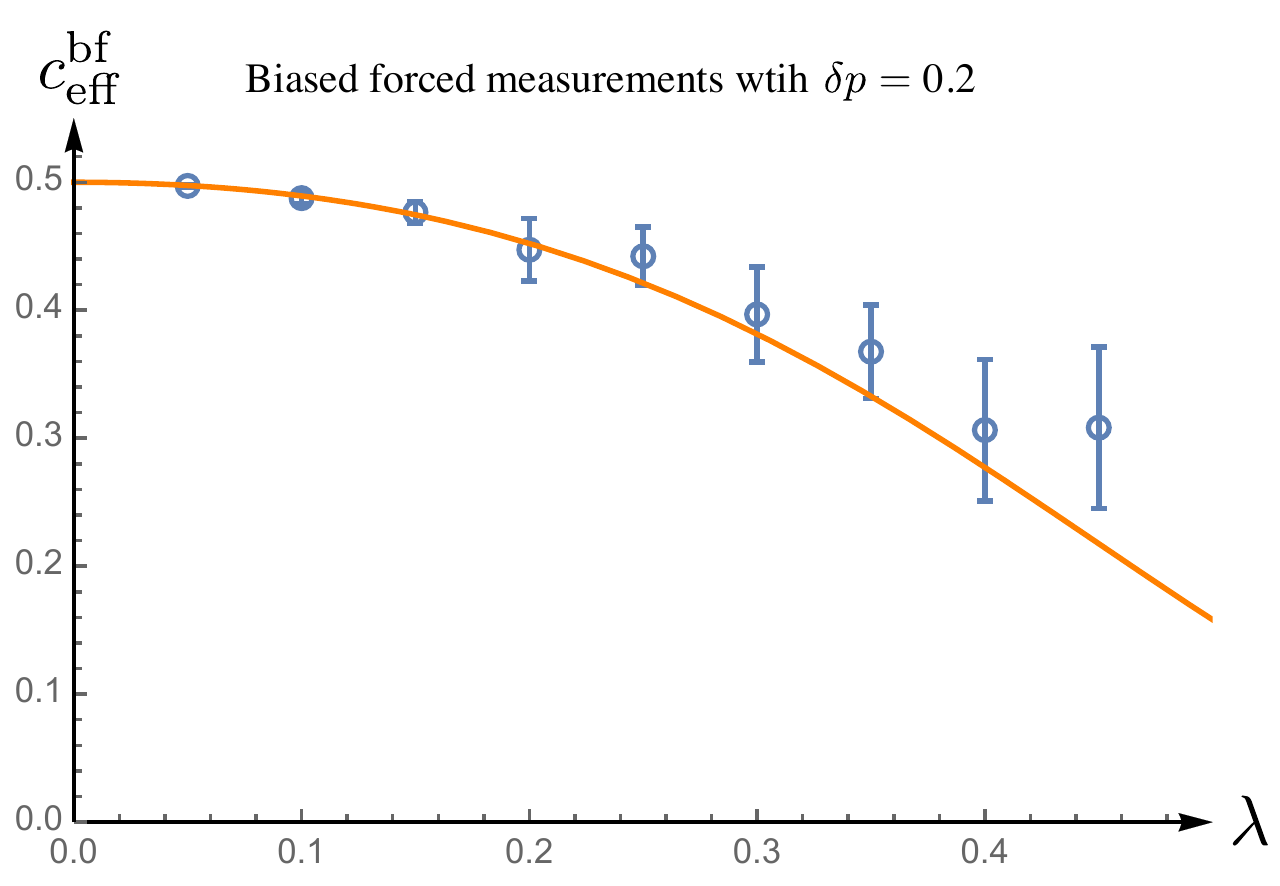}}\\
    \subfloat[]{\includegraphics[width=.4\linewidth]{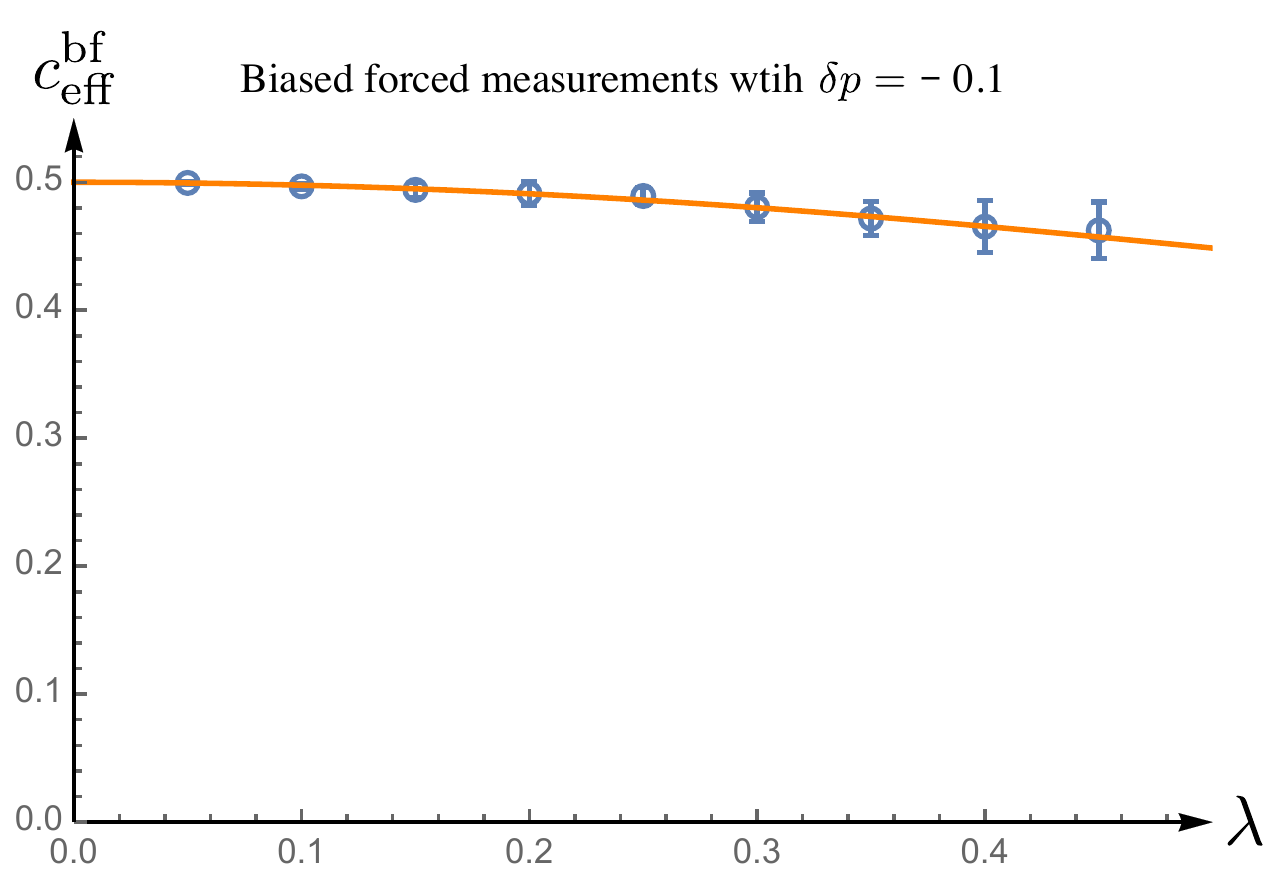}}
    \hspace{.05\linewidth}
    \subfloat[]{\includegraphics[width=.4\linewidth]{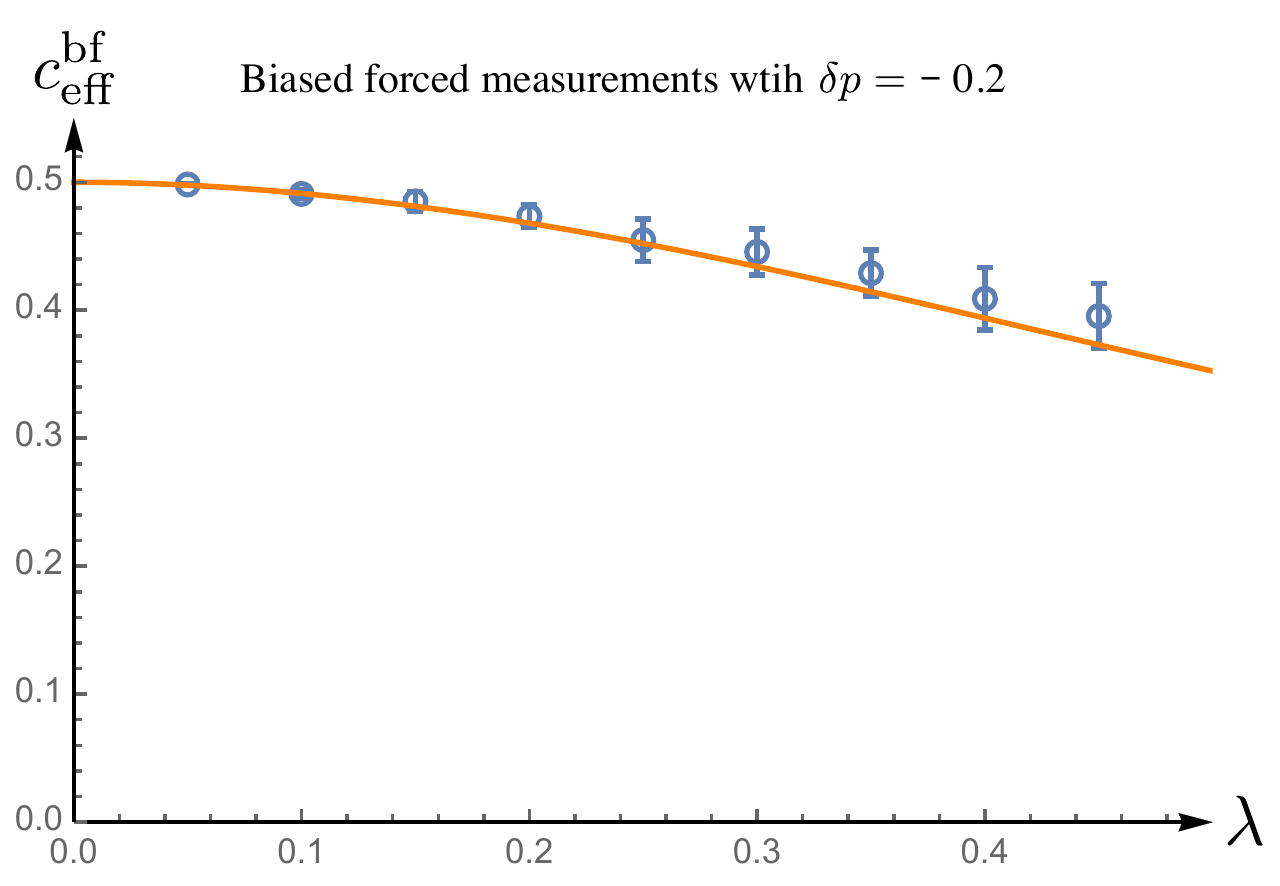}}
    \caption{The effective central charge $\Ceff^{\rm bf}$ for the biased forced measurements with $p_+ = p_b^x+\delta p$. The orange curves are given by Eq. \eqref{eq:ceff_bforced}. The blue circles are the effective central charge obtained from MPS simulations.}
    \label{fig:ceff_bias}
\end{figure*}

In Fig. \ref{fig:ceff_bias}, we compare our analytical result Eq. \eqref{eq:ceff_bforced} with the numerical simulations and find good agreements between them (especially at small measurement strength $\lambda$). 

\bibliography{ref}
\end{document}